\documentclass[aps,prc,floatfix,showpacs,twocolumn,nofootinbib]{revtex4-1}
\usepackage{ulem}
\usepackage{dcolumn}
\usepackage{bm}
\usepackage{color}
\usepackage{amssymb}
\usepackage{amsmath}
\usepackage{graphicx}
\usepackage{amsfonts}
\usepackage{slashed}
\usepackage{pstricks}
\usepackage{float}
\usepackage{hyperref}
\usepackage{array}
\allowdisplaybreaks
\usepackage{dcolumn}
\usepackage{epsf}
\usepackage{float}
\usepackage{afterpage}

\usepackage{tikz}
\usepackage[style=base]{caption}
\usepackage[style=base]{subcaption}
\usepackage{multirow}
\usepackage{dcolumn} % for aligning tables based on decimals
\newcolumntype{d}[1]{D{.}{.}{#1}} 
\newcommand\Tstrut{\rule{0pt}{2.6ex}} % = Top strut for tables

\newcommand{\Train}{\mathbb{T}} % generic training set (really a set of indices)
\newcommand{\Test}{\mathbb{O}} % generic testing/'out of sample' set (really a set of indices)
\newcommand{\Valid}{\mathbb{V}} % generic validation set (really a set of indices)

\usepackage[ruled,vlined]{algorithm2e}

\newcommand{\vect}[1]{\mathbf{#1}}

\begin{document}
\title{Uncertainty Quantification-Enabled Inversion of Nuclear Euclidean Responses}
\author{
{Krishnan} Raghavan$^{\, {\rm a} }$,
%ORCID: 0000-0001-9409-2011
{Alessandro} Lovato$^{\, {\rm b,c,d} }$,
}
\affiliation{
$^{\,{\rm a}}$\mbox{Mathematics and Computer Science Division, Argonne National Laboratory, Lemont, Illinois 60439, USA}\\
$^{\,{\rm b}}$\mbox{Physics Division, Argonne National Laboratory, Lemont, Illinois 60439, USA}\\
$^{\,{\rm c}}$\mbox{Computational Science Division, Argonne National Laboratory, Lemont, Illinois 60439, USA}\\
$^{\,{\rm d}}$\mbox{INFN-TIFPA Trento Institute of Fundamental Physics and Applications, Via Sommarive, 14, 38123 {Trento}, Italy}
}
\date{\today}
%%%%
\begin{abstract} 
Nuclear quantum many-body methods rely on integral transform techniques to infer properties of electroweak response functions from ground-state expectation values. Retrieving the energy dependence of these responses is highly non-trivial, especially for quantum Monte Carlo methods, as it requires inverting the Laplace transform -- a notoriously ill-posed problem. In this work, we propose an artificial neural network architecture suitable for accurate response function reconstruction with precise estimation of the uncertainty of the inversion. We demonstrate the capabilities of this new architecture benchmarking it against Maximum Entropy and previously developed neural network methods designed for a similar task, paying particular attention to its robustness against increasing noise in the input Euclidean responses.  
\end{abstract}
%\pacs{24.10.Cn,25.30.Pt,26.60.-c}
%%%%%
\maketitle

\section{Introduction}
The combination of sophisticated nuclear forces systematically derived within effective theories of QCD and numerical methods solving the quantum many-body problem with high accuracy~\cite{Barrett:2013nh,Hagen:2013nca,Hergert:2015awm,Carbone:2013eqa,Epelbaum:2011md,Carlson:2014vla} has enabled ab-initio studies of the structure of several nuclides across the nuclear chart, including $^{208}$Pb~\cite{Hu:2021trw}. Although existing many-body methods can describe nuclear ground-state properties and low-energy electroweak transitions with high accuracy~\cite{Gysbers:2019uyb}, modeling real-time nuclear dynamics still poses a key challenge for current computational methods. Accessing it is crucial for achieving a fully microscopic understanding of processes such as fission, heavy-ion fusion, as well as lepton- and nucleus-nucleus scattering. Generally, computing dynamical properties of quantum many-body systems remains one of the paradigmatic open problems in quantum many-body theory, primarily due to quantum interference~\cite{Roggero:2018hrn}. Computational limitations often strongly constrain the physical regimes in which quantum many-body dynamics can be solved on classical computers. Emerging technologies, such as neural-network quantum states~\cite{carleo:2017,Schmitt:2020} and quantum computing~\cite{Miessen:2023}, hold great promise in this area, but their applications to nuclear physics are still in their infancy.

In this work, we will focus on the linear response regime, whose applications are ubiquitous in physics — including neutron scattering on materials and photon scattering in atomic systems — and specifically on lepton-nucleus scattering. A quantitative description of the latter is critical for the interpretation of inclusive and semi-exclusive electron-nucleus scattering experiments, shedding light on short-range correlations and the transition between hadronic and partonic degrees of freedom~\cite{Hen:2016kwk,Segarra:2019gbp,Segarra:2020plg}. Additionally, the success of the accelerator neutrino program hinges on precise theoretical calculations of neutrino-nucleus scattering, as they are essential for reconstructing the oscillated flux from measurements of particles produced in the aftermath of the scattering process~\cite{Benhar:2015wva,Katori:2016yel,NuSTEC:2017hzk}.

State-of-the-art nuclear many-body methods, such as Green's function Monte Carlo (GFMC) and Coupled-Cluster, derive information about electroweak response functions from their integral transforms, which can be expressed as ground-state expectation values~\cite{Lovato:2016gkq,Lovato:2020kba, Sobczyk:2021dwm} ---  a notable exception in this regard consists in using an appropriate expansion in Chebyshev polynomials~\cite{Sobczyk:2021ejs}. However, reconstructing the energy dependence of these response functions presents non-trivial challenges, especially when utilizing the Laplace kernel, as in the GFMC~\cite{Carlson:2001mp}. 

The maximum entropy method (MaxEnt)~\cite{Bryan:1990,Jarrell:1996}, widely employed in condensed matter and lattice-QCD applications, has proven accurate in inverting the Laplace transform and reconstructing smooth response functions, characterized by a single broad quasi-elastic peak. On the other hand, MaxEnt struggles in the low-energy region, which is often characterized by several peaks, corresponding to low-energy nuclear transitions. For this reason, to retrieve the electromagnetic response of $^{12}$C these transitions had to be removed from the Euclidean response, using available experimental data~\cite{Lovato:2016gkq}. Such shortcomings also yield certain discrepancies between GFMC and exact Faddeev results for the $^{3}$H muon capture rate near the nuclear breakup threshold, corresponding to energies of a few MeV~\cite{Lovato:2019fiw}

Inspired by earlier machine-learning applications~\cite{McCann:2017,Arsenault:2017,Yoon:2018,Xie:2019,Fournier:2020}, in Ref.~\cite{Raghavan:2020bze} a ``Physics informed'' artificial neural network (Phys-NN) was introduced for approximating the inverse of the Laplace transform. Phys-NN has proven to outperform MaxEnt in both the low-energy transfer and the quasielastic regions, and to be more robust against noise in the input Euclidean responses. However, similarly to MaxEnt, Phys-NN is not able not propagate the statistical uncertainties of the Euclidean response into the response function and to quantify the systematic errors due to the approximate inversion of the Laplace transform.

In this work, we overcome this limitation by developing an artificial neural network architecture that provides accurate response functions with quantified uncertainties, dubbed ``UQ-NN''. To achieve this goal, we capitalize on a flexible parametrization of the response functions inspired by the one used in MaxEnt, which guarantees fast convergence of the training phase. As a result, UQ-NN exhibits an improved accuracy of the inversion and increased robustness to noise as compared to Phys-NN.

The present manuscript is structured as follows. In Sec.~\ref{sec:responses} we state the problem to be solved and discuss the relevant features of the nuclear electromagnetic responses. In Sec.~\ref{sec:ML} we describe our artificial-neural network architecture. In Sec.~\ref{sec:res} we present our results, and in Sec.~\ref{sec:conc} we discuss our conclusions. 
\section{Electroweak responses from their Laplace transforms}
\label{sec:responses}
The nuclear response functions relevant to describing inclusive lepton-nucleus scattering cross sections in the one-boson exchange approximation can be generically written as:
\begin{align}
R({\bf q},\omega) &= \sum_f \left\langle 0 | j^\dagger({\bf q},\omega) |f\right\rangle \left\langle f | j({\bf q},\omega) |0\right\rangle \nonumber \\
& \times \delta(E_f-\omega-E_0),
\label{eq:res_def}
\end{align}
In the above equation, $|0\rangle$ and $|f\rangle$ are the initial and final nuclear states with energies $E_0$ and $E_f$, respectively, and $j({\bf q},\omega)$ denotes the electroweak current operators.

In order to avoid computing all transitions induced by the current operator --- which is impractical except for very light nuclear systems~\cite{Shen:2012xz,Golak:2018qya} --- the GFMC infers properties of the response functions from their Laplace transform~\cite{Carlson:1992ga}, which is defined as:
\begin{equation}
E_\alpha({\bf q},\tau) = \int_0^\infty d\omega, e^{-\omega\tau} R_\alpha({\bf q},\omega) , .
\label{eq:euc_def}
\end{equation}

Fixing the intrinsic energy dependence of the charge and current operators to the quasi-elastic (QE) peak, $\omega_{\rm QE}=\sqrt{{\bf q}^2+m^2}-m$, where $m$ denotes the mass of the nucleon, one can express the Euclidean responses as ground-state expectation values:
\begin{equation*}
E_\alpha({\bf q},\tau)=\langle 0 | j_\alpha^\dagger ({\bf q},\omega_{\rm QE}) e^{-(H-E_0)\tau} j({\bf q},\omega_{\rm QE}) | 0\rangle,
\end{equation*}
where $H$ is the nuclear Hamiltonian. These expectation values can be evaluated by using the GFMC method on a uniform grid of $n_\tau$ imaginary-time points~\cite{Carlson:1992ga,Carlson:2001mp}. Standard GFMC calculations entail $n_\tau = 150$, with a maximum imaginary time of $0.07$ MeV$^{-1}$. A set of noisy estimates for $E_\alpha({\bf q},\tau_i)$ can be obtained by performing independent imaginary-time propagations, from which the average Euclidean response $\bar{E}_\alpha({\bf q},\tau_i)$, and their associated statistical errors $\sigma_i$ can be readily estimated. 

In addition to the imaginary time, the continuous variables $\omega$ is also discretized on $n_\omega$ grid points, so that Eq.~\eqref{eq:euc_def} becomes
\begin{equation}
E_i=\sum_{j=1}^{n_\omega} K_{ij} R_j
\label{eq:euc_discrete}
\end{equation}
where $K_{ij}=e^{-\omega_j\tau_i}\Delta\omega_j$ and $R_j\equiv R(\omega_j)$. The GFMC responses functions are typically tabulated on a grid of $n_\omega = 2000$ points with a maximum of $2$ GeV. The log-likelihood of the reconstructed responses is proportional to  
\begin{equation}
\chi^2[\mathbf{R}, \bar{\mathbf{E}}]=\frac{1}{n_{\tau}}\sum_i \frac{(\sum_j K_{ij} R_j - \bar{E}_i)^2}{\sigma_i^2}\, .
\label{eq:chi_squared}
\end{equation}
where $\bar{\mathbf{E}} \in \mathbb{R}^{n_\tau}$, $\mathbf{R} \in \mathbb{R}^{n_\omega}$. Note that, GFMC calculations provide the sample covariance matrix between the data at $\tau=\tau_i$ and $\tau=\tau_j$, which is typically non-diagonal because of correlations among the imaginary-time points~\cite{Carlson:1992ga,Carlson:2001mp,Lovato:2016gkq}. However, the likelihood reduces to the one of Eq.~\ref{eq:chi_squared} once the data and the Laplace kernel are rotated in the basis where the covariance is a diagonal~\cite{Jarrell:1996}. 

Due to the smoothing effect of the Laplace kernel, a simple minimization of $\chi^2$ results in multiple response functions that are consistent, within errors, with the GFMC Euclidean response. Maximum entropy methods~\cite{Bryan:1990, Jarrell:1996} aim to address this ambiguity by treating the response functions, both positive definite and normalizable, as probability distributions. According to the the principle of maximum entropy, their values are determined by maximizing the entropy, defined as
\begin{equation}
S[\mathbf{R}, \mathbf{M}] = \sum_{i=1}^{n_\omega} \left[ R_i - M_i - R_i\ln\left(\frac{R_i}{M_i}\right) \right]\Delta\omega_i\, ,
\label{eq:entropy}
\end{equation}
The positive valued function $M(\omega)$ serves as the default model and incorporates prior knowledge about $R(\omega)$ when no data is available and $\mathbf{M} \in \mathbb{R}^{n_\omega}$. The entropy quantifies the deviation between the response function and the model. It has a maximum value of zero when $\mathbf{R} = \mathbf{M}$ and it is negative otherwise.

By applying Bayes' theorem, MaxEnt identifies the most probable response function, minimizing the quantity
\begin{equation}
Q[\mathbf{R}, \bar{\mathbf{E}}, \mathbf{M}] = \frac{1}{2}\chi^2[\mathbf{R}, \bar{\mathbf{E}}] - \alpha S[\mathbf{R}, \mathbf{M}]\, ,
\label{eq:q_def}
\end{equation}
with respect to $\mathbf{R}$. Here $\alpha$ is a parameter that determines the balance between the entropy and the log-likelihood. When $\alpha=0$, the standard ill-posed minimization is recovered, and for $\alpha \gg 1$, $\mathbf{R}$ converges to the default $\mathbf{M}$.
In our study, we carry out all benchmarks against the historic MaxEnt approach~\cite{Gull:1978}, which selects $\alpha$ such that $\chi^2 = 1$. More sophisticated methods like the {\it classic} MaxEnt~\cite{Skilling:1989} and {\it Bryan} MaxEnt~\cite{Bryan:1990} tend to overfit the data~\cite{VonDerLinden:1999,Hohenadler:2005}. The arbitrariness in choosing $\alpha$ hampers a robust reconstruction of $R(\omega)$ in the small-$\omega$ region. Specifically, too small $\alpha$ results in overfitting $E(\tau)$ and uncontrolled oscillations in the reconstructed responses.

The key point in the inversion of the Laplace transform resides in the minimization of $Q[\mathbf{R},\bar{\mathbf{E}}, \mathbf{M}]$, defined in Eq.~\ref{eq:q_def} for given $\bar{\mathbf{E}}$ and $\mathbf{M}$. An efficient way to accomplish this task was first discussed in Ref.~\cite{Bryan:1990}, and it entails performing a singular value decomposition (SVD) of the kernel 
\begin{equation}
K=V \Sigma U^T\, .
\end{equation}
In the above equation, $U$ and $V$ are $n_\omega\times n_\omega$  and $n_\tau\times n_\tau$ orthogonal 
matrices, while $\Sigma$ is a $n_\tau\times n_\omega$ rectangular diagonal matrix. Since the kernel is effectively singular, the smallest elements on the diagonal are essentially zero for the numerical precision. Hence, without loss of accuracy, we keep only the $n_s$ largest eigenvalues and disregard the others so that only the first $n_s$ columns of $U$ are relevant for representing the kernel.

The gradient of the log-likelihood is given by 
\begin{equation}
\frac{\partial \chi^2}{\partial R_i}=\sum_j \frac{\partial \chi^2 }{ \partial E_j} \frac{\partial E_j}{\partial R_i}=K^{T}_{ij} \frac{\partial \chi^2 }{ \partial E_j}\,.
\end{equation}
Since the columns of $K^T$ are linear combinations of the ones of $U$, all the search directions for the minimum are spanned, within machine precision, by the first $n_s$ columns of $U$. In this {\it singular space}, the stationary condition of $Q[R]$ reads
\begin{align}
0 = \frac{\partial Q}{\partial R_i} & = \alpha \frac{\partial S}{\partial R_i} - \frac{1}{2}\frac{\partial \chi^2}{\partial R_i}=0\, ,
\end{align}
which implies
\begin{align}
-\alpha \ln(R_i/M_i)=\frac{1}{2}\sum_j K^{T}_{ij}\frac{\partial \chi^2}{\partial E_j}.
\label{eq:bryan_min}
\end{align}
Thus, the solution can be represented in terms of the vector $\vect{u}$
\begin{equation}\label{eq:int_E}
\ln\Big(\frac{R_i}{m_i}\Big)=K^{T}_{ij}u_j.
\end{equation}
Since only the first $n_s$ elements of $\Sigma$ are different from zero, not all the components of $\vect{u}$ are independent. Since $K^T$ and $U$ share the same vector space and since most of the relevant search directions
lie in the singular space, the solution can be written in the form
\begin{equation}\label{eq:Bryan}
R_i=M_i \exp\Big (\sum_{j=1}^{n_s} U_{ij} u_j \Big)\, .
\end{equation}
Therefore, to the machine-precision level, the most general solution of Eq. (\ref{eq:bryan_min}) only depends on the $n_s$ coordinates $u_j$. In MaxEnt applications, owing to the ranges of $\omega$ and $\tau$ in GFMC calculations, $n_s \simeq 30 \ll n_\omega=2000$. Hence, a standard Newton procedure to minimize $Q[\mathbf{R},\bar{\mathbf{E}}, \mathbf{M}]$ converges much faster for finding $u_j$ than for the original $R_i$.

\begin{figure*}[!ht]
    \centering
    \includegraphics[width = 0.85\textwidth]{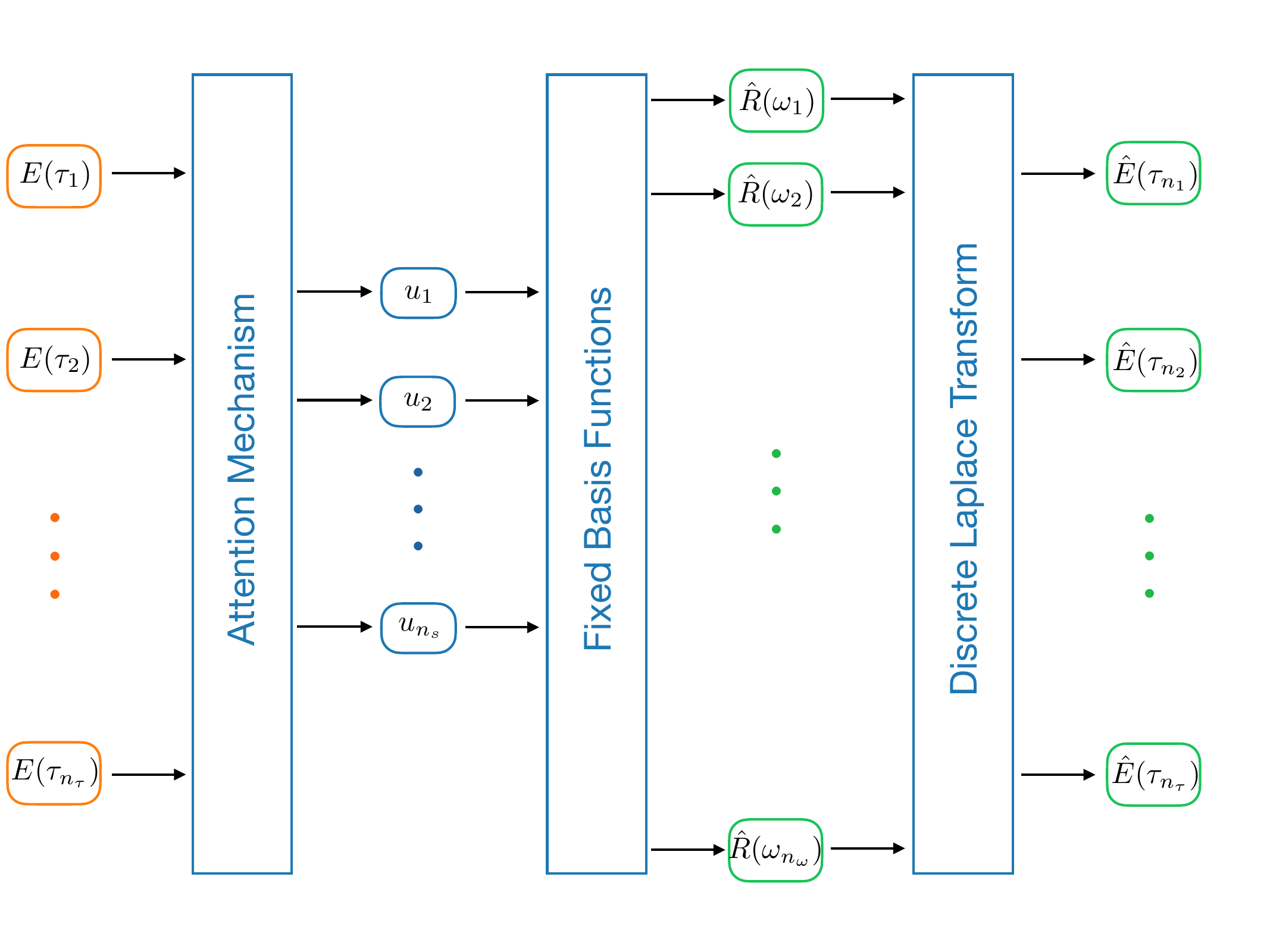}
    \caption{Schematic overview of the Ent-NN architecture.}
    \label{fig:ent_nn}
\end{figure*}

\section{Artificial neural network inversion algorithm} 
\label{sec:ML}
In a recent work~\cite{Raghavan:2020bze}, a physics-informed artificial neural network (Phys-NN) was introduced to approximating the inverse of the Laplace transform. Phys-NN employs a Gaussian kernel basis to capture its structure of the Laplace kernel. In this work we utilize instead the more advantageous parameterization of Eq.~\eqref{eq:Bryan}, using artificial neural-networks to determine the coefficients $u_j$. Formally, the reconstructed response is given by
\begin{equation}
\hat{R}_i(\boldsymbol{\theta})= m_i \exp \Big (\sum_{j=1}^{n_s} U_{ij} u_j(\boldsymbol{\theta}) \Big)\ ,
\label{eq:bryan}
\end{equation}
where $\boldsymbol{\theta}$ denotes the collection of training parameters. This entails a critical reduction of the artificial neural network output dimension compared to Phys-NN, whose outputs where directly the $n_\omega$ values $\hat{R}_i$. 

\subsection{Entropy Neural Network~(Ent-NN)}
The first architecture discussed in this work, dubbed ``Ent-NN'', takes as input the $n_\tau$ discrete Euclidean response values and provides the corresponding response functions.
\begin{equation} \label{eq:EntNN}
u_j(\boldsymbol{\theta})=f(\mathbf{E}; \boldsymbol{\theta}).
\end{equation}

The architecture of Ent-NN, displayed in Fig.~\ref{fig:ent_nn}, is comprised of three central elements: i) the attention mechanism comprised of two feed-forward layers with one skip connection that takes as input $E_i$ and generates the coefficients $u_i(\boldsymbol{\theta})$, ii) the fixed basis function $U_{ij}$, used to estimate $\hat{\mathbf{R}}(\boldsymbol{\theta})$ from Eq.~\eqref{eq:bryan}, and iii) the discrete Laplace transform of Eq.~\eqref{eq:euc_discrete} for computing the Euclidean $\hat{\mathbf{E}}(\boldsymbol{\theta})$ associated to the reconstructed response function as $ \hat{E}_i(\boldsymbol{\theta})=\sum_{j=1}^{n_\omega} K_{ij} \hat{R}_j(\boldsymbol{\theta}).$

\subsubsection{Training}
As in Ref.~\cite{Raghavan:2020bze}, Ent-NN is trained on two distinct datasets comprising pairs of physically relevant $R(\omega)$, $E(\tau)$. The responses belonging to the first dataset are characterized by a single broad asymmetric peak, corresponding to the QE reaction mechanism, modeled by a skew-normal distribution. The responses belonging to the second dataset exhibit a sharper elastic (EL) peak at low energy, in addition to the QE one. The corresponding Euclidean responses are obtained by applying the discrete Laplace transform of Eq.~\eqref{eq:euc_discrete}. Since the simulated responses are smooth functions of $\omega$, the numerical integration error on the Euclidean responses is about $10^{-5}.$ For each of the one-peak and two-peaks cases, we generate a total of $500,000$ pairs $(\mathbf{R}_k,\mathbf{E}_k)\in \mathbb{R}^{n_\omega + n_\tau}$ of responses and corresponding Euclidean responses, which we then partition into training~($\Train$), validation~($\Valid$), and test/out-of-sample~($\Test$) datasets. The one-peak and two-peak test datasets comprise $1,000$ pairs each; the combined test dataset is just the union of these two sets. We use 80\% and 20\% of the remaining data for training the network and validation, respectively. 

The optimal values for the parameters $\boldsymbol{\theta}$ are found by the standard supervised learning approach of approximately solving
\begin{equation}
\label{eq:batched_loss}
	     \min_{\boldsymbol{\theta}}  \, \frac{1}{|\Train|} \sum_{k\in \Train} \ell\left(\mathbf{E}_k, \mathbf{R}_k, \hat{\mathbf{R}}_k(\boldsymbol{\theta})\right)
\end{equation}
by using a minibatch-based stochastic gradient descent procedure to minimize an empirical loss function. Our overall objective in the above equation is the average loss over the $|\Train|$ points in the training set. Taking inspiration from MaxEnt, for each data and model output, we employ a loss function that is the sum of a \textit{response} and a \textit{Euclidean cost}
\begin{align*} 
&\ell(\mathbf{E}_k, \mathbf{R}_k, \hat{\mathbf{R}}_k(\boldsymbol{\theta})) = \nonumber\\
&\qquad \gamma_{R} |S(\mathbf{R}_k, \hat{\mathbf{R}}_k(\boldsymbol{\theta}))| +  \gamma_E \chi^2(\mathbf{E}_k,\hat{\mathbf{R}}_k(\boldsymbol{\theta}))\,.
\end{align*}
As discussed below, the positive-definite constants  $\gamma_{R}$ and $\gamma_{E}$ are chosen to compensate for the fact that $\chi^2(\mathbf{E}_k,\hat{\mathbf{R}}_k(\boldsymbol{\theta}))$ is typically much larger than the entropy $S_R(\mathbf{R}_k, \hat{\mathbf{R}}_k(\boldsymbol{\theta}))$. The response cost --- closely related to the Kullback–Leibler divergence~\cite{kullback1951information} --- ensures that the reconstructed response functions are close to the original ones. The absolute value ensures that the response cost has a minimum value of $0$ when $\mathbf{R}_k = \hat{\mathbf{R}}_k(\boldsymbol{\theta})$ and is positive otherwise. The Euclidean cost is aimed at aligning the Laplace transform $\hat{\mathbf{E}}(\boldsymbol{\theta})$ of the reconstructed response functions with the original Euclidean responses. 

Since the inversion of the Laplace transform is an ill-posed problem, there are many response functions, possibly wildly different among each other, whose Laplace transform are compatible with the original Euclidean responses within statistical uncertainties. Consequently, there are instances in which $\chi^2_E$ is small even when the reconstructed response is not similar to the original one, leading to potential instabilities in the minimization procedure. To tame this behavior, we  split the training into two phases.

\begin{figure}[!ht]
    \centering
    \includegraphics[width=\linewidth]{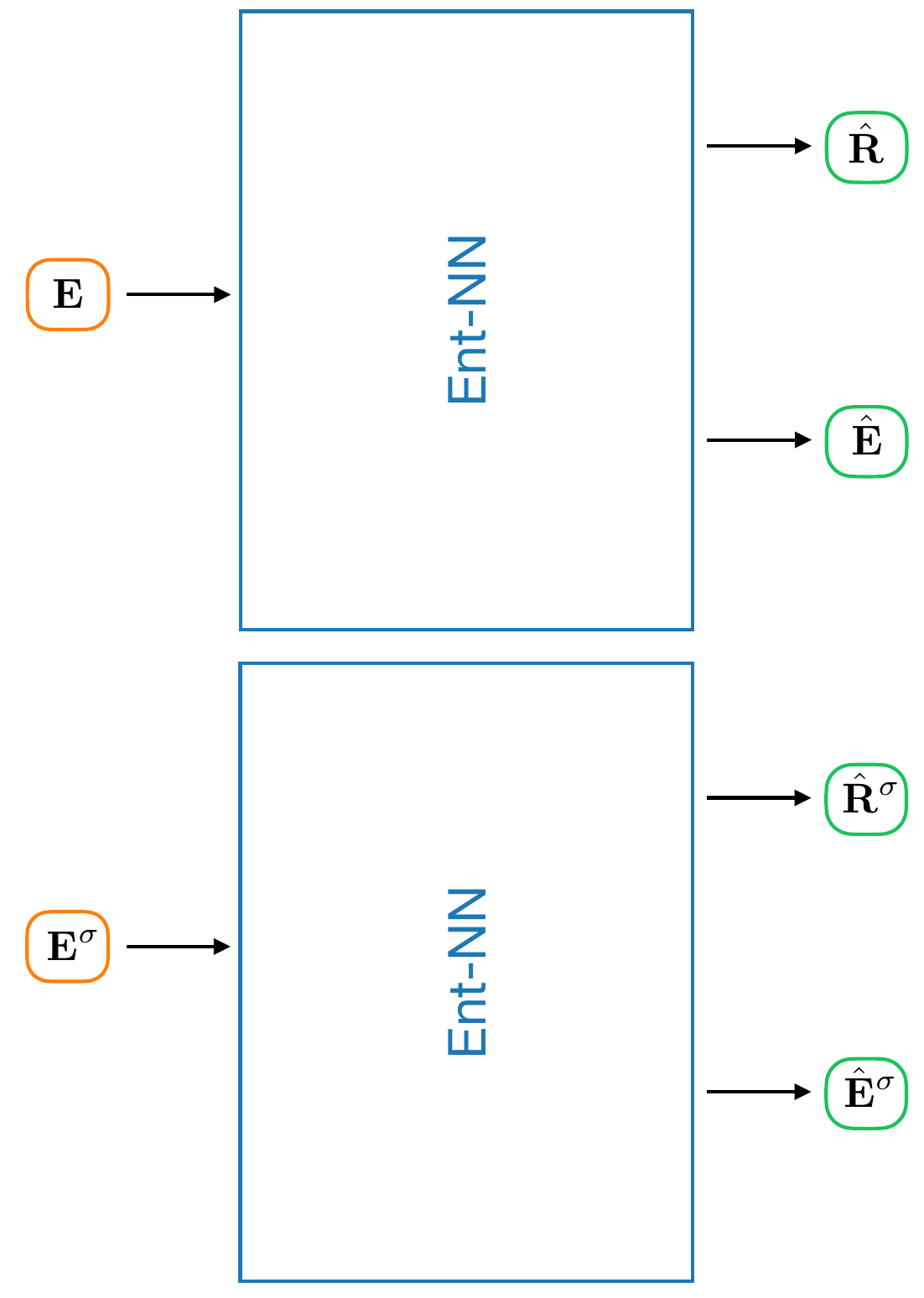}
  \caption{Schematic overview of the UQ-NN architecture.} \label{fig:UQ-NN}
\end{figure}

In the first phase, we take $\gamma_{R} = 10^7$ and $\gamma_E = 10^{-7}$ and optimize the network using the Adam~\cite{kingma2014adam} optimizer with a learning rate of $10^{-4}$. Since $\gamma_R \gg \gamma_E$, the entropy response cost dominates the loss function and drives the reconstructed response functions close to the original ones. Once the entropy cost has reduced significantly, we enter the second phase of the optimization, where we keep $\gamma_{R} = 10^{7}$ while increasing the relative importance of the Euclidean cost by taking $\gamma_E = 1$ so that Ent-NN also learns to keep the Laplace transform of the response function close to the original Euclidean response. Reducing the learning rate in the second phase to $10^{-5}$ is necessary in order to keep the reconstructed response functions close to the optimal ones found in the previous phase.

\subsection{Uncertainty Quantification Neural Network~(UQ-NN)}
Meaningful comparisons between GFMC calculations of the response function with experimental data require carrying out rigorous uncertainty quantification. The latter is particularly relevant when making predictions for neutrino-nucleus scattering, as cross-section uncertainties should be carefully propagated in the error-budget of neutrino-oscillation parameters~\cite{Ruso:2022qes}. 

Approximately inverting the Laplace transform using artificial neural-network entails two distinct sources of uncertainty. The first one is due to the choice of the neural network model, which includes the set of optimal parameters found in the training procedure. In Ref.~\cite{Raghavan:2020bze}, this effect was found to be small. The second one concerns propagating the statistical errors associated with GFMC estimates of $\mathbf{E}$ into the reconstructed responses. 

The distribution of the computed Euclidean response computed within the GFMC is Gaussian, hence
\begin{equation}
P(\mathbf{E}) \propto \exp\left(-\frac{(\mathbf{E} - \bar{\mathbf{E}})^2}{2\sigma^2}\right)\,.
\label{eq:gauss_E}
\end{equation}
Consistent with the notation of Sec.~\ref{sec:responses}, we assume a diagonal covariance matrix and, to simplify the discussion, we also consider the standard deviation $\sigma$ to be independent of $\tau$ — both assumptions can be easily relaxed. The corresponding probability distribution of the response functions is then given by 
\begin{equation}
P(\mathbf{R}) = \int d\mathbf{E} P(\mathbf{R} | \mathbf{E}) P(\mathbf{E})
\label{eq:propagation}
\end{equation}
Assuming that the response functions can be accurately reconstructed using artificial neural networks and that the training parameters are narrowly distributed around the optimal ones implies that
\begin{equation}
P(\mathbf{R} | \mathbf{E}) = \delta( \mathbf{R} - \hat{\mathbf{R}}(\boldsymbol{\theta})) 
\end{equation}
Hence, following standard Monte Carlo error propagation procedures~\cite{Zhang:2021}, samples of response functions distributed according to $P(\mathbf{R})$ of Eq.~\eqref{eq:propagation} can, in principle, be obtained by drawing Euclidean responses from the Gaussian distribution of Eq.~\eqref{eq:gauss_E} and applying Ent-NN to each of them. 

The first task is accomplished by adding stochastic noise~\cite{Fournier:2020} to the simulated Euclidean responses as
\begin{equation}
E_i^\sigma = \bar{E}_i + \epsilon_i\, ,
\label{eq:eq_DG}
\end{equation}
where $\epsilon_i$ are independent samples from a Gaussian distribution with mean zero and standard deviation $\sigma$~\cite{Raghavan:2020bze}. We note that including an energy-dependent error, $\sigma_i$, would involve drawing $\epsilon_i$ from Gaussian distributions characterized by energy-dependent widths. 

As for the second step above, applying Ent-NN to the Euclidean obtained as in Eq.~\eqref{eq:eq_DG} results in exceedingly large variations in the reconstructed responses. The reason for this behavior has to be ascribed to the fact that adding Gaussian noise to the Euclidean responses makes them significantly different from the ones found in the original training dataset. Hence, Ent-NN is forced to extrapolate, leading to inaccurate reconstructions, as apparent by the significant discrepancies between the reconstructed Euclidean responses $\hat{\mathbf{E}}(\boldsymbol{\theta})$ and the noisy ones $\mathbf{E}^\sigma $ that are taken as input. 

To remedy this shortcoming, we have developed UQ-NN. Its architecture, schematically illustrated in Fig.~\ref{fig:UQ-NN}, comprises two Ent-NN neural networks operating in parallel. The first takes as input the original $\mathbf{E}$ and returns the corresponding $\hat{\mathbf{R}}(\boldsymbol{\theta})$ as well as $\hat{E}_i(\boldsymbol{\theta})=\sum_{j=1}^{n_\omega} K_{ij} \hat{R}_j(\boldsymbol{\theta})$. The second takes as input the noisy $\mathbf{E}^\sigma$ and returns $\hat{\mathbf{R}}^\sigma(\boldsymbol{\theta})$ and $\hat{E}_i^\sigma(\boldsymbol{\theta})=\sum_{j=1}^{n_\omega} K_{ij} \hat{R}^\sigma_j(\boldsymbol{\theta})$. Crucially, the training parameters are common to the two Ent-NN blocks.

%-------------------------------------------------------------------------------------
\subsubsection{Training}
The training process for UQ-NN follows the steps as in Ent-NN, with the distinction that each batch of data also includes the noisy Euclidean responses of Eq.~\eqref{eq:eq_DG}. Hence, the optimal values of $\boldsymbol{\theta}$ are found by 
\begin{equation}
\label{eq:batched_loss_uq}
	     \min_{\boldsymbol{\theta}}  \, \frac{1}{|\Train|} \sum_{k\in \Train} \ell\left(\mathbf{E}_k, \mathbf{R}_k, \mathbf{E}^\sigma_k,\hat{\mathbf{R}}_k(\boldsymbol{\theta}), \hat{\mathbf{R}}^\sigma_k(\boldsymbol{\theta})\right)
\end{equation}
Crucially, no noisy response appears among the arguments of the loss function, as there is no direct way to generate it from $\mathbf{E}^\sigma_k$. The loss function is a generalization of the one of Ent-NN 
\begin{align} 
& \ell\left(\mathbf{E}_k, \mathbf{R}_k, \mathbf{E}^\sigma_k,\hat{\mathbf{R}}_k(\boldsymbol{\theta}), \hat{\mathbf{R}}^\sigma_k(\boldsymbol{\theta})\right) = \gamma_{R} |S(\mathbf{R}_k, \hat{\mathbf{R}}_k(\boldsymbol{\theta}))| \nonumber\\
& \qquad \qquad +  \gamma_E [\chi^2(\mathbf{E}_k,\hat{\mathbf{R}}_k(\boldsymbol{\theta})) + \chi^2(\mathbf{E}^\sigma_k,\hat{\mathbf{R}}^\sigma_k(\boldsymbol{\theta}))]\,.
\label{eq:cost_UQNN}
\end{align}
The additional term in the Euclidean cost drives the reconstructed $\mathbf{E}^\sigma(\boldsymbol{\theta})$ to be close to the input noisy Euclidean, thereby providing a reliable reconstruction of the corresponding noisy response function. Specifically, for each Euclidean response $\mathbf{E}_k$ in a batch of data, we also sample $\mathbf{E}^\sigma$ as in Eq.~\eqref{eq:eq_DG}. We then employ UQ-NN to evaluate the corresponding $\hat{\mathbf{R}}_k(\boldsymbol{\theta})$ and $\hat{\mathbf{R}}^\sigma(\boldsymbol{\theta})$. Finally, we numerically integrate them to obtain the corresponding Euclidean, which are used in the cost function defined in Eq.~\eqref{eq:cost_UQNN}. The rest of the hyperparameters and the choice of $\gamma_R$ and $\gamma_E$ are identical to the training of Ent-NN.

The training is performed for different noise levels. The smallest is $\sigma = 10^{-5}$, which is similar to the statistical noise of actual GFMC calculations of $^4$He Euclidean electromagnetic responses. We also consider $\sigma = 10^{-4}$, $\sigma=10^{-3}$ --- a value compatible with typical GFMC calculations of $^{12}$C --- and $\sigma=10^{-2}$. This latter noise level corresponds to typical auxiliary-field diffusion Monte Carlo~\cite{Schmidt:1999lik} (AFDMC) calculation of $^{16}$O that are currently being performed.

\begin{table}[!htb]
\caption{Ent-NN, UQ-NN, Phys-NN, and MaxEnt testing metrics $\overline{S_R}$, and $\overline{\chi_E^{2}}$ for the one-peak, two-peak, and combined datasets.}
\label{tab:results}
\begin{tabular}{l||c|c}
            &     $\overline{\chi^{2}_E}$ & $\overline{S_R} \times 10^{-4}$ \\[3pt]
            \hline \hline 
% & \multicolumn{2}{c}{\textbf{Ent-NN}} \Tstrut \\[3pt]
% One-peak    & 3.5(3.49)     &  0.49(56) \\ 
% Two-peak    & 5.05(147.00)     &  1.06(17)  \\ 
% Combined    & 3.535(121.00)        &  0.732(47)   \\[3pt]
% \hline 
%             & \multicolumn{2}{c}{\textbf{UQ-NN}} \Tstrut \\[3pt]
% One-peak    & 6.20(9.77)    & .43(48)  \\ 
% Two-peak    & 10.76(553.00)    & 1.25(16) \\ 
% Combined    & 12.35(612)         & 0.859(33)      \\[3pt]
% \hline 
%             & \multicolumn{2}{c}{\textbf{Phys-NN}} \Tstrut \\[3pt]
% One-peak    & 2.00(1.41)    & 0.62(55)  \\ 
% Two-peak    & 7.76(19.72)   & 1.44(23)  \\ 
% Combined    & 5.15(22)      & 1.03(17)  \\[3pt]
% \hline 
& \multicolumn{2}{c}{\textbf{Ent-NN}} \Tstrut \\[3pt]
One-peak    & 3.594        &  0.492 \\ 
Two-peak    & 5.053      &  1.068 \\ 
Combined    & 3.535      &  0.732   \\[3pt]
\hline 
            & \multicolumn{2}{c}{\textbf{UQ-NN}} \Tstrut \\[3pt]
One-peak    & 6.202    & 0.436  \\ 
Two-peak    & 10.750  & 1.255 \\ 
Combined    & 12.351   & 0.859      \\[3pt]
\hline
            & \multicolumn{2}{c}{\textbf{Phys-NN}} \Tstrut \\
One-peak    & 2.002       & 0.622 \\ 
Two-peak    & 7.766      & 1.442  \\ 
Combined    & 5.153       & 1.031 \\
            & \multicolumn{2}{c}{\textbf{MaxEnt}} \Tstrut \\[3pt]
One-peak    & 1.015    & 60.4  \\ 
Two-peak    & 1.016    & 107  \\ 
Combined    & 1.015    & 83.7  
\end{tabular}
\end{table}

% %%%%%%%%%%%%%%%%% Box plots
\begin{figure*}[!htb]
\begin{subfigure}{0.45\textwidth}
    \includegraphics[width = \columnwidth]{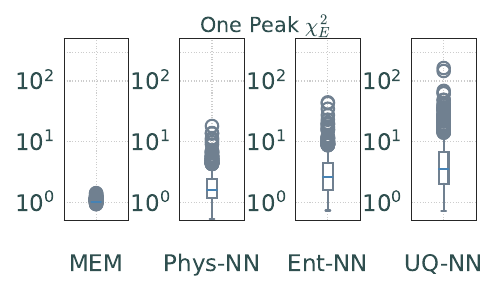}
\end{subfigure}%%
\begin{subfigure}{0.45\textwidth}
    \includegraphics[width = \columnwidth]{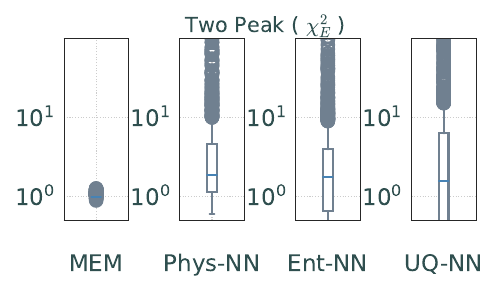}
\end{subfigure}
\begin{subfigure}{0.45\textwidth}
    \includegraphics[width = \columnwidth]{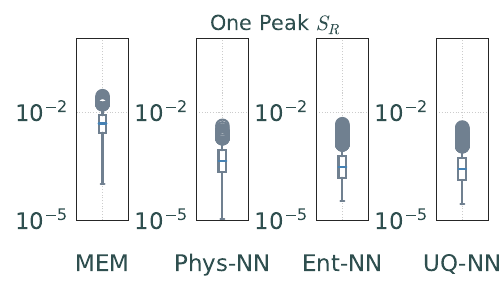}
\end{subfigure} %%
\begin{subfigure}{0.45\textwidth}
    \includegraphics[width = \columnwidth]{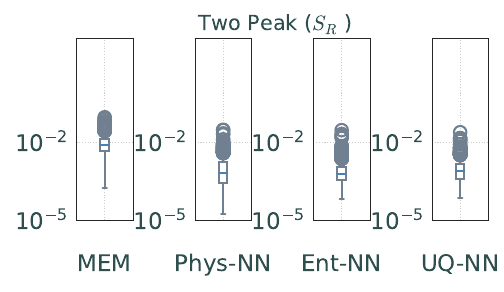}
\end{subfigure}
\caption{Box plots of (top row) $\chi_E^{2}$, and (bottom row) $S_R$ for the one peak dataset~(left) and two peak dataset~(right) as obtained with Ent-NN, UQ-NN, Phys-NN, and MaxEnt. The line in the middle of the box denotes the median, and the box represents the range between the 25\% and 75\% quantiles. Whiskers cover the area between the 1\% and 99\% quantiles; data beyond these whiskers are outliers and are indicated by circles.}
\label{fig:results_box}
\end{figure*}

\section{Results}
\label{sec:res}
\subsection{Model Performance}
To quantify the accuracy of both Ent-NN and UQ-NN, we adopt two metrics averaged over the test/out-of-sample dataset $\Test$. The first one is the average absolute value of the entropy
\begin{equation*}
\overline{S_R} = \frac{1}{|\Test|} \sum_{k\in \Test} |S_R(\mathbf{R}_k, \hat{\mathbf{R}}_k(\boldsymbol{\theta}))|\,,  \end{equation*}
where the entropy functional is defined in Eq.~\eqref{eq:entropy}.  It is important to note that a smaller $\overline{S_R}$ corresponds to more accurate reconstructed responses. The second one is the average reduced $\chi^2_{E}$
\begin{equation*}
\overline{\chi^2_{E}} = \frac{1}{|\Test|} \sum_{k\in \Test} \chi^2(\mathbf{E}_k, \hat{\mathbf{R}}_k(\boldsymbol{\theta}))\,,
\end{equation*}
where the log-likelihood is the one of Eq.~\eqref{eq:chi_squared}. 

Table~\ref{tab:results} summarizes the testing metrics for the single-peak, two-peak, and combined datasets, comparing Ent-NN and UQ-NN against Phys-NN and MaxEnt. Note that, while the MaxEnt metrics are identical to those found in Ref.~\cite{Raghavan:2020bze}, Phys-NN exhibits lower entropy and a larger $\chi^2$. This change in performance is a deliberate choice in the training process. To better reconstruct the responses, we found it beneficial to use a larger $\gamma_E$ and a smaller $\gamma_R$ in the loss function. In this initial comparison, no noise has been added to the input Euclidean responses, which only suffer from the numerical integration error discussed in Sec.~\ref{sec:ML}. All approaches perform best in reconstructing one-peak responses, while the accuracy of two-peak reconstructions appears to be lower. The reconstructions for the combined dataset fall between those of the other two datasets. This behavior is expected, considering that response functions characterized by two peaks, especially with the EL peak located in the low-$\omega$ region, are notoriously more complicated to reconstruct than those with a single broad QE peak.

Regarding the entropy metric, Ent-NN and UQ-NN significantly outperform Phys-NN and MaxEnt across the one-peak, two-peak, and combined datasets. The improved performance of Ent-NN and UQ-NN in capturing the energy dependence of the response functions compared to Phys-NN is a direct result of utilizing the basis functions outlined in Eq.~\eqref{eq:bryan}, which are tailored for inverting the Laplace transform. Conversely, historic MaxEnt yields the least accurate reconstructions. Ent-NN and Phys-NN generate similar reconstruction values, but Ent-NN has a slight edge in terms of entropy. Although UQ-NN's reconstructions are less precise than Ent-NN's, they outperform both Phys-NN and MaxEnt.

In comparing the $\chi^2$ values, MaxEnt appears to perform the best. However, this is due to the historical MaxEnt method, where the optimal response function is determined by setting $\alpha$ in Eq.\eqref{eq:q_def} to achieve $\chi^2_E=1$. As indicated by the entropy metric, the inherently ill-posed nature of the problem implies that a $\chi^2_E\approx 1$ does not guarantee an accurate reconstruction of the original response functions. In fact, despite Ent-NN, UQ-NN, and Phys-NN having higher $\chi^2$ values than MaxEnt, they yield more accurate response functions, as evidenced by the entropy values. It is important to note that Ent-NN provides a smaller $\chi^2$ value than Phys-NN, while UQ-NN is slightly less accurate. This behavior can be attributed mainly to the training process of UQ-NN, which involves introducing noise into the Euclidean responses, as discussed in Section\ref{sec:ML}. As expected, introducing noise into the model inherently degrades its accuracy. On the other hand, it enhances the model's robustness, allowing it to propagate the uncertainties of the Euclidean responses in the reconstructed response functions.

% %%%%%%%%%%############## COrrelation plots
\begin{figure}[!b]
\begin{subfigure}{0.48\linewidth}
    \includegraphics[keepaspectratio,width = \columnwidth]{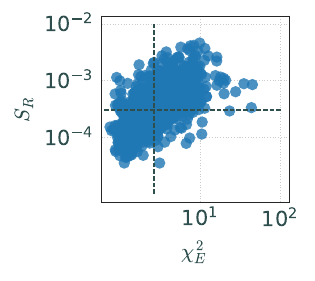}
\end{subfigure}
\begin{subfigure}{0.48\linewidth}
    \includegraphics[keepaspectratio,width = \columnwidth]{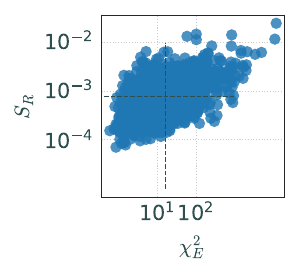}
\end{subfigure}\\
\begin{subfigure}{0.45\linewidth}
    \includegraphics[keepaspectratio,width = \columnwidth]{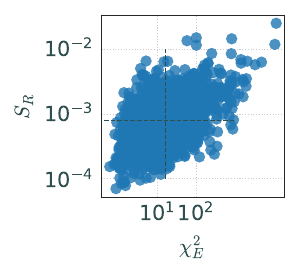}
\end{subfigure}
\begin{subfigure}{0.45\linewidth}
    \includegraphics[keepaspectratio,width = \columnwidth]{corr_chi2_Entropy_UQ_two.pdf}
\end{subfigure}
\caption{Ent-NN (left column) and UQ-NN (right colums) scatter plots of $\chi^2_{E}$ versus $S_R$. The top (bottom) row refers the one-peak (two-peaks) dataset. The dashed lines indicate the median $\overline{\chi^2_{E}}$ and $\overline{S_R}$ values.}
\label{fig:corr}
\end{figure}

\begin{figure*}[!htb]
\includegraphics[width = \textwidth]{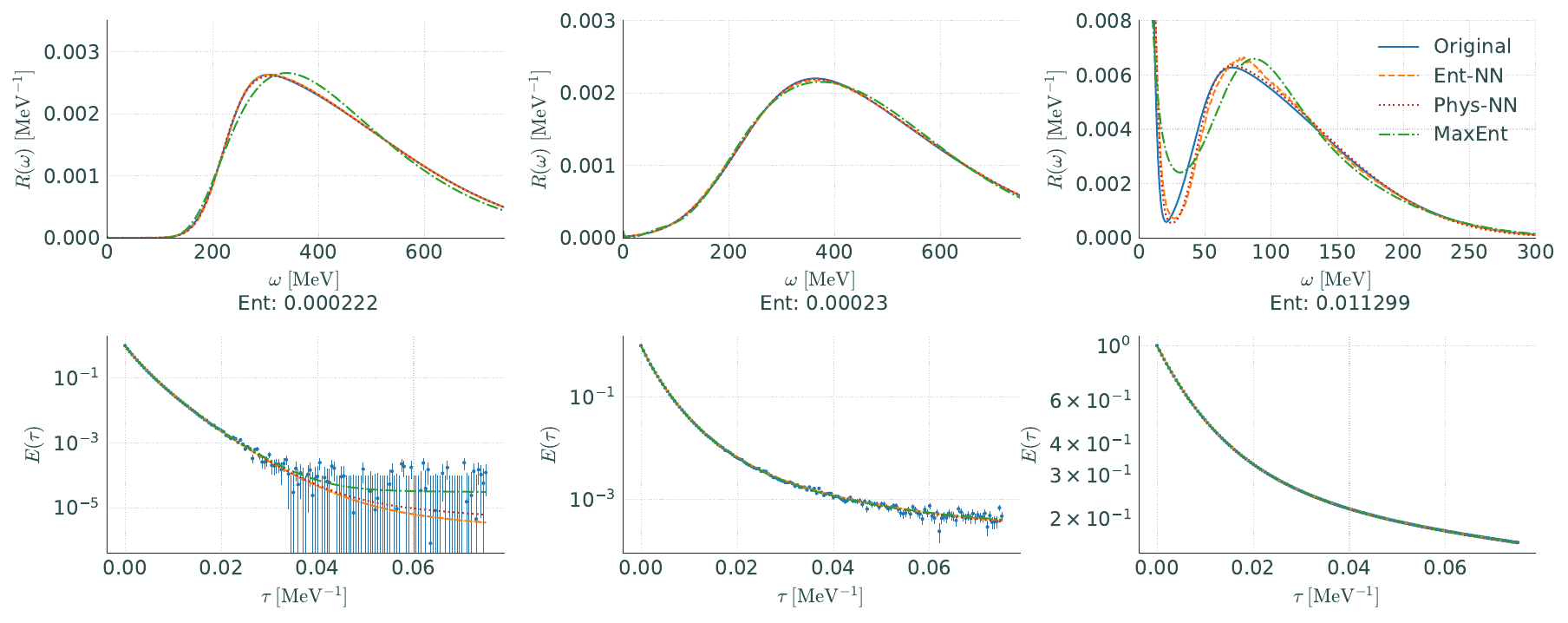}
\caption{Comparison between the Ent-NN, Phys-NN, and MaxEnt reconstructions for the combined dataset. The top row displays the response functions and the bottom row the corresponding Euclidean responses.}
\label{fig:results_recon}
\end{figure*}

Fig.~\ref{fig:results_box} displays the box plot of the $S_R$ and $\chi_E^{2}$ distributions for the one-peak(left column) and two-peak~(right column) datasets obtained within the Ent-NN, UQ-NN, Phys-NN, and MaxEnt methods. Consistent with the results listed in Table~\ref{tab:results}, the one-peak $\chi_E^{2}$ and $S_R$ distributions are narrower and centered on smaller values than the two-peak ones, while the combined dataset results are intermediate between the two. Since Ent-NN, UQ-NN, and Phys-NN are trained to keep the reconstructed response function close to the original ones, we observe a much smaller spread of $\overline{S_R}$ values compared with MaxEnt.

As previously discussed, historic MaxEnt naturally produces $\chi^2_E$ values tightly clustered around one. In contrast, the spread associated with Ent-NN and UQ-NN is more extensive, as evident in the scatter plots of Fig.\ref{fig:corr}. Correlations between the $\chi^{2}_E$ values and $S_R$ are noticeable in both neural network architectures, particularly for the two-peak dataset. This correlation provides a crucial advantage over MaxEnt, serving as a tool to assess the accuracy of the Laplace transform inversion. For instance, the outliers in $\chi^{2}_E$ visible in the top right corners of all panels in Fig.\ref{fig:corr} serve as clear indicators of imperfect reconstructions of the response functions.

\begin{figure*}[!htb]
 \includegraphics[ width = 0.99\textwidth]{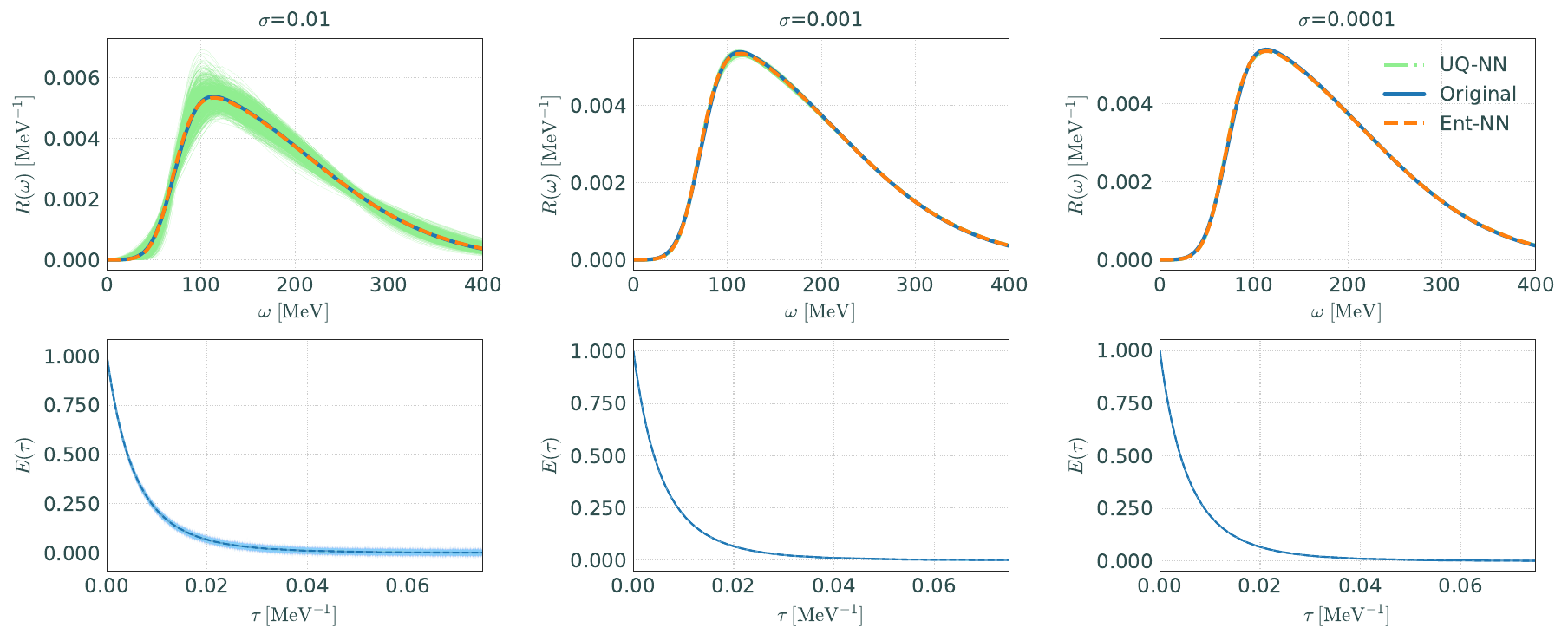}
\caption{Top row: one-peak responses obtained via the UQ-NN architecture (green band), compared to Ent-NN (dashed orange line) and the original response (blue solid line). Bottom row: corresponding Euclidean responses with varying noise level.}
 \label{fig:results_recon_one}
\end{figure*} %%

 \begin{figure*}[!htb]
            \includegraphics[width = 0.99\textwidth]{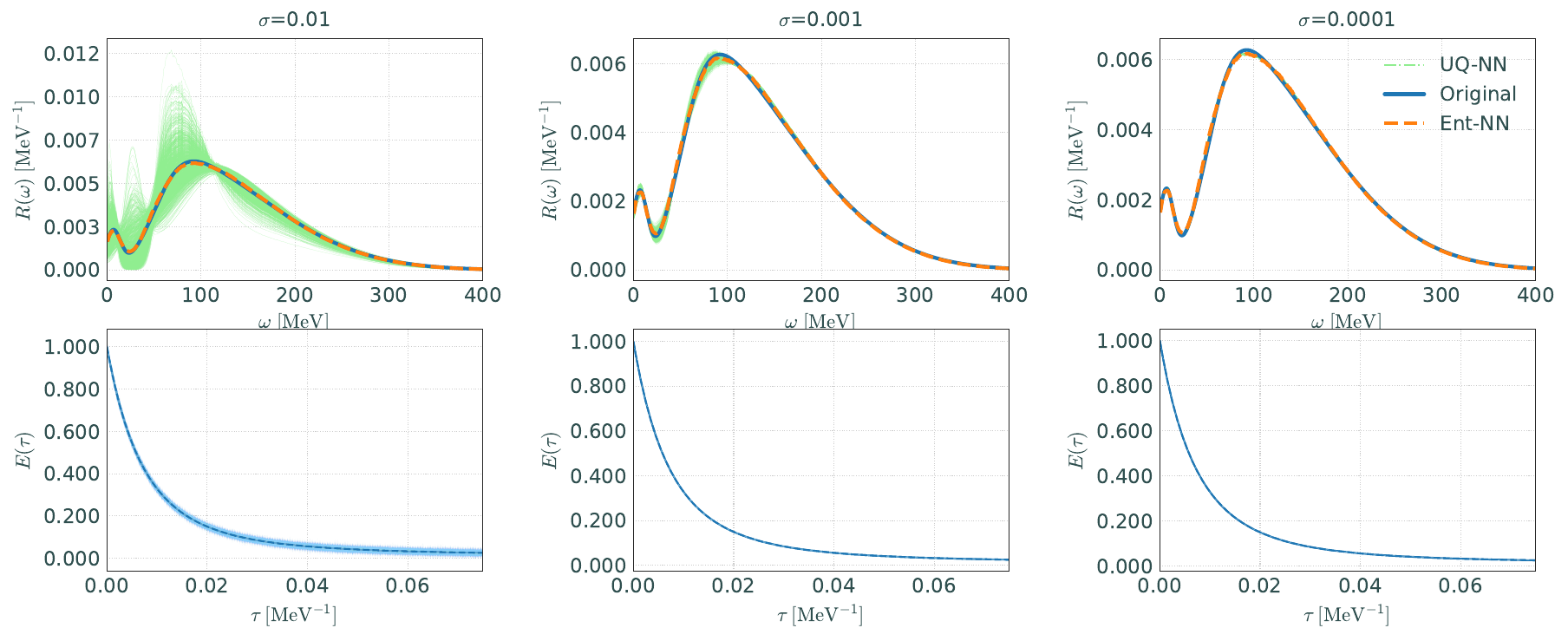}
        \caption{Same as Fig.~\ref{fig:results_recon_one} for the two-peaks dataset.}
            \label{fig:results_recon_two}
    \end{figure*}

Fig.~\ref{fig:results_recon} provides a comparative illustration of the various inversion approaches. From the combined dataset, we select the \textit{best} (left panels), \textit{average} (central panels), and \textit{worst} (right panels) reconstructed response functions based on their $S_R$ values obtained within Ent-NN. Remarkably, not only the ``best'' and the ``average'' response functions but also the ``worst'' response functions reconstructed with Ent-NN and UQ-NN exhibit closer agreement with the original ones compared to those obtained with historic MaxEnt. We also note that Phys-NN performs similarly to Ent-NN and UQ-NN. Additionally, the Laplace transforms of the Ent-NN and UQ-NN  response functions show excellent alignment with the original Euclidean responses: the $\chi^2_E$ values are $0.37$, $8.15$, and $342.27$ for the best, average, and worst reconstructions, respectively.

\subsection{Uncertainty Quantification}
As discussed in Sec.~\ref{sec:ML}, to perform uncertainty quantification, we generated one-thousand Euclidean responses according to Eq.~\eqref{eq:eq_DG} for both the one- and two-peaks datasets. The corresponding reconstructed responses, distributed according to $P(\mathbf{R})$ of Eq.~\eqref{eq:propagation}, are obtained applying UQ-NN to each of these Euclidean responses. Both the input noisy Euclidean and reconstructed responses are displayed in Figure~\ref{fig:results_recon_one} and Figure~\ref{fig:results_recon_two} for the one- and two-peaks datasets, respectively, with varying degrees of noise in the input. The latter ranges from $\sigma=10^{-4}$, corresponding to GFMC calculations of $^{12}$C, to  $\sigma=10^{-2}$, which is appropriate for ongoing AFDMC calculations of $^{16}$O Euclidean responses. Note that in the inference phase, as opposed to the training, we do not input the original Euclidean to UQ-NN, but just the noisy one, which is the only one available in real-world scenarios. 

The key feature of our UQ-NN model is that it is able to capture the uncertainties present in the input data. In fact, the uncertainty in the reconstructed responses is generally proportional to the amount of statistical noise in the input Euclidean. This is already apparent in the one-peak dataset results; the green band around the reconstructed responses becomes larger as the noise level in the input increases. Notably, the original response function remains always enveloped within these bands, corroborating the accuracy of UQ-NN in both reconstructing the response function and in propagating the uncertainties of the input Euclidean. 

Similar observations can be made for the two-peaks dataset. Here, however, we can observe some additional features of the UQ-NN responses. First, at the largest noise level $\sigma=10^{-2}$, a three-peaks structure seems to emerge in the low-$\omega$ region, despite no response functions in the training dataset have more than two peaks. We ascribe the origin of this rich structure to the noise added to the Euclidean, which may yield a three-peak structure in some of the reconstructed $\hat{\mathbf{R}}(\boldsymbol{\theta})$. Secondly, for the lowest noise level, $\sigma=10^{-4}$, UQ-NN fails to precisely capture the $\omega$ dependence of the original response function, even in the QE peak region. We checked that Ent-NN (and even Phys-NN) suffers from similar limitations. One possible reason for this behavior are numerical errors associated with numerically computing the Laplace transform --- see Eq.~\eqref{eq:euc_discrete} --- when generating the training data set. The latter could be larger than the estimated $10^{-5}$ value, especially for responses with two peaks. Another possibility is the uncertainty inherent to the neural-network model, which includes the set of optimal parameters found in the training procedure and the training itself. To better estimate the latter, we plan on using deep Bayesian Neural Network~\cite{Utama:2015hva,Niu:2018csp}, which in the context on Nuclear Physics, have proven reliable in predicting masses and radii of several nuclei across the nuclear chart, with quantified uncertainties.

\section{Conclusions}
\label{sec:conc}
Microscopic calculations of nuclear electromagnetic response functions are essential for connecting aspects of short- and long-range nuclear dynamics observed in electron-scattering experiments~\cite{Donnelly:1975ze,Benhar:2006wy,Hen:2016kwk,Kegel:2021jrh,PREX:2021umo,CREX:2022kgg} to the underlying nuclear interactions and currents. Additionally, the availability of accurate electroweak (neutral and charge-current) response functions with quantified theoretical uncertainties is crucial for the success of the accelerator neutrino program~\cite{Benhar:2015wva,Katori:2016yel,NuSTEC:2017hzk}, as nuclear cross section uncertainties are among the primary sources of systematic errors.

Over the past decade, the GFMC method has been extensively employed to compute electroweak response functions of nuclei with up to $A=12$ nucleons, including one and two-body current operators consistent with the Hamiltonian generating correlations in the initial and final state of the reactions~\cite{Lovato:2016gkq,Lovato:2020kba,Andreoli:2021cxo}. More recently, Coupled Cluster theory has achieved remarkable success in modeling longitudinal and transverse electromagnetic responses of nuclei as large as $^{40}$Ca~\cite{Sobczyk:2020qtw,Sobczyk:2021dwm,Sobczyk:2023sxh}, retaining one-body current contributions only. In contrast to methods relying on harmonic-oscillator expansions~\cite{Sobczyk:2021dwm}, GFMC faces no challenges in handling high-resolution (or high-momentum) nuclear forces. These capabilities are indispensable for modeling the final state of reactions with momentum transfers above $q\simeq 400$ MeV. However, reconstructing electroweak response functions from GFMC imaginary-time propagators involves solving the notoriously ill-posed problem of inverting the Laplace transform.

In this study, we introduced two artificial neural network architectures suitable for approximating the inversion of the Laplace transform: Ent-NN and UQ-NN. A significant advantage over existing architectures, such as Phys-NN~\cite{Raghavan:2020bze}, is that both Ent-NN and UQ-NN utilize basis functions tailored to the Laplace kernel, determined through its singular value decomposition~\cite{Bryan:1990}. We demonstrated their effectiveness by benchmarking Ent-NN and UQ-NN against Phys-NN and MaxEnt, using a substantial dataset comprising synthetic yet realistic data characterized by a broad quasielastic peak and a sharper elastic peak at lower energies. Ent-NN outperforms both Phys-NN and MaxEnt in terms of both metrics we considered. As a significant advance with respect to existing approaches, UQ-NN is designed to propagate the statistical errors of the Euclidean response through the response functions, which is critical for carrying out quantitative comparisons with experimental data.

The results presented in this work are particularly relevant for extending quantum Monte Carlo calculations to nuclei larger than $^{12}$C, specifically to ongoing AFDMC calculations of electroweak response functions of $^{16}$O. Standard MaxEnt suffers from severe limitations in this case for two main reasons. First, AFDMC suffers from significantly larger statistical noise than GFMC, primarily due to a stronger fermion sign problem~\cite{Schmidt:1999lik}. Second, the electromagnetic responses of $^{16}$O exhibit a rich low-energy structure, especially in the longitudinal channel, including elastic transitions and collective modes~\cite{Bacca:2014rta}.

Notably, while our architectures were explicitly developed to approximate the inverse of the Laplace transform, they can be readily extended to different kernels, including the Lorentz one. Consequently, they serve as valuable benchmarks for other inversion techniques, such as those based on expanding response functions on regularized ansatz~\cite{Efros:1999ab}, which require imposing the breakup threshold or employing appropriate expansions in Chebyshev polynomials~\cite{Sobczyk:2021ejs}.

\acknowledgments
We express our sincere gratitude to R.~B.~Wiringa and C.~Adams for their careful reading of the manuscript and their valuable suggestions.  We are deeply grateful for the enlightening and stimulating discussions with P.~Balaprakash, N.~Rocco, and S.~Wild, whose valuable insights significantly contributed to the development of this work. Additionally, we express our sincere thanks to G.~ Carleo for his unwavering support for this research direction. This work was supported in part by the U.S.\ Department of Energy (DOE), Office of Science, Offices of Advanced Scientific Computing Research and Nuclear Physics, by the Argonne LDRD program, and by the NUCLEI, FASTMath, and RAPIDS SciDAC projects under contract number DE-AC02-06CH11357. A.L., is also supported by DOE Early Career Research Program awards. We are grateful for the computing resources from the Joint Laboratory for System Evaluation  and Leadership Computing Facility at Argonne.

\section*{Bibliography}

\bibliography{biblio}

\begin{thebibliography}{56}%
\makeatletter
\providecommand \@ifxundefined [1]{%
 \@ifx{#1\undefined}
}%
\providecommand \@ifnum [1]{%
 \ifnum #1\expandafter \@firstoftwo
 \else \expandafter \@secondoftwo
 \fi
}%
\providecommand \@ifx [1]{%
 \ifx #1\expandafter \@firstoftwo
 \else \expandafter \@secondoftwo
 \fi
}%
\providecommand \natexlab [1]{#1}%
\providecommand \enquote  [1]{``#1''}%
\providecommand \bibnamefont  [1]{#1}%
\providecommand \bibfnamefont [1]{#1}%
\providecommand \citenamefont [1]{#1}%
\providecommand \href@noop [0]{\@secondoftwo}%
\providecommand \href [0]{\begingroup \@sanitize@url \@href}%
\providecommand \@href[1]{\@@startlink{#1}\@@href}%
\providecommand \@@href[1]{\endgroup#1\@@endlink}%
\providecommand \@sanitize@url [0]{\catcode `\\12\catcode `\$12\catcode
  `\&12\catcode `\#12\catcode `\^12\catcode `\_12\catcode `\%12\relax}%
\providecommand \@@startlink[1]{}%
\providecommand \@@endlink[0]{}%
\providecommand \url  [0]{\begingroup\@sanitize@url \@url }%
\providecommand \@url [1]{\endgroup\@href {#1}{\urlprefix }}%
\providecommand \urlprefix  [0]{URL }%
\providecommand \Eprint [0]{\href }%
\providecommand \doibase [0]{http://dx.doi.org/}%
\providecommand \selectlanguage [0]{\@gobble}%
\providecommand \bibinfo  [0]{\@secondoftwo}%
\providecommand \bibfield  [0]{\@secondoftwo}%
\providecommand \translation [1]{[#1]}%
\providecommand \BibitemOpen [0]{}%
\providecommand \bibitemStop [0]{}%
\providecommand \bibitemNoStop [0]{.\EOS\space}%
\providecommand \EOS [0]{\spacefactor3000\relax}%
\providecommand \BibitemShut  [1]{\csname bibitem#1\endcsname}%
\let\auto@bib@innerbib\@empty
%</preamble>
\bibitem [{\citenamefont {Barrett}\ \emph {et~al.}(2013)\citenamefont
  {Barrett}, \citenamefont {Navratil},\ and\ \citenamefont
  {Vary}}]{Barrett:2013nh}%
  \BibitemOpen
  \bibfield  {author} {\bibinfo {author} {\bibfnamefont {B.~R.}\ \bibnamefont
  {Barrett}}, \bibinfo {author} {\bibfnamefont {P.}~\bibnamefont {Navratil}}, \
  and\ \bibinfo {author} {\bibfnamefont {J.~P.}\ \bibnamefont {Vary}},\ }\href
  {\doibase 10.1016/j.ppnp.2012.10.003} {\bibfield  {journal} {\bibinfo
  {journal} {Prog. Part. Nucl. Phys.}\ }\textbf {\bibinfo {volume} {69}},\
  \bibinfo {pages} {131} (\bibinfo {year} {2013})}\BibitemShut {NoStop}%
\bibitem [{\citenamefont {Hagen}\ \emph {et~al.}(2014)\citenamefont {Hagen},
  \citenamefont {Papenbrock}, \citenamefont {Hjorth-Jensen},\ and\
  \citenamefont {Dean}}]{Hagen:2013nca}%
  \BibitemOpen
  \bibfield  {author} {\bibinfo {author} {\bibfnamefont {G.}~\bibnamefont
  {Hagen}}, \bibinfo {author} {\bibfnamefont {T.}~\bibnamefont {Papenbrock}},
  \bibinfo {author} {\bibfnamefont {M.}~\bibnamefont {Hjorth-Jensen}}, \ and\
  \bibinfo {author} {\bibfnamefont {D.~J.}\ \bibnamefont {Dean}},\ }\href
  {\doibase 10.1088/0034-4885/77/9/096302} {\bibfield  {journal} {\bibinfo
  {journal} {Rept. Prog. Phys.}\ }\textbf {\bibinfo {volume} {77}},\ \bibinfo
  {pages} {096302} (\bibinfo {year} {2014})},\ \Eprint
  {http://arxiv.org/abs/1312.7872} {arXiv:1312.7872 [nucl-th]} \BibitemShut
  {NoStop}%
\bibitem [{\citenamefont {Hergert}\ \emph {et~al.}(2016)\citenamefont
  {Hergert}, \citenamefont {Bogner}, \citenamefont {Morris}, \citenamefont
  {Schwenk},\ and\ \citenamefont {Tsukiyama}}]{Hergert:2015awm}%
  \BibitemOpen
  \bibfield  {author} {\bibinfo {author} {\bibfnamefont {H.}~\bibnamefont
  {Hergert}}, \bibinfo {author} {\bibfnamefont {S.~K.}\ \bibnamefont {Bogner}},
  \bibinfo {author} {\bibfnamefont {T.~D.}\ \bibnamefont {Morris}}, \bibinfo
  {author} {\bibfnamefont {A.}~\bibnamefont {Schwenk}}, \ and\ \bibinfo
  {author} {\bibfnamefont {K.}~\bibnamefont {Tsukiyama}},\ }\href {\doibase
  10.1016/j.physrep.2015.12.007} {\bibfield  {journal} {\bibinfo  {journal}
  {Phys. Rept.}\ }\textbf {\bibinfo {volume} {621}},\ \bibinfo {pages} {165}
  (\bibinfo {year} {2016})},\ \Eprint {http://arxiv.org/abs/1512.06956}
  {arXiv:1512.06956 [nucl-th]} \BibitemShut {NoStop}%
\bibitem [{\citenamefont {Carbone}\ \emph {et~al.}(2013)\citenamefont
  {Carbone}, \citenamefont {Cipollone}, \citenamefont {Barbieri}, \citenamefont
  {Rios},\ and\ \citenamefont {Polls}}]{Carbone:2013eqa}%
  \BibitemOpen
  \bibfield  {author} {\bibinfo {author} {\bibfnamefont {A.}~\bibnamefont
  {Carbone}}, \bibinfo {author} {\bibfnamefont {A.}~\bibnamefont {Cipollone}},
  \bibinfo {author} {\bibfnamefont {C.}~\bibnamefont {Barbieri}}, \bibinfo
  {author} {\bibfnamefont {A.}~\bibnamefont {Rios}}, \ and\ \bibinfo {author}
  {\bibfnamefont {A.}~\bibnamefont {Polls}},\ }\href {\doibase
  10.1103/PhysRevC.88.054326} {\bibfield  {journal} {\bibinfo  {journal} {Phys.
  Rev. C}\ }\textbf {\bibinfo {volume} {88}},\ \bibinfo {pages} {054326}
  (\bibinfo {year} {2013})},\ \Eprint {http://arxiv.org/abs/1310.3688}
  {arXiv:1310.3688 [nucl-th]} \BibitemShut {NoStop}%
\bibitem [{\citenamefont {Epelbaum}\ \emph {et~al.}(2011)\citenamefont
  {Epelbaum}, \citenamefont {Krebs}, \citenamefont {Lee},\ and\ \citenamefont
  {Meissner}}]{Epelbaum:2011md}%
  \BibitemOpen
  \bibfield  {author} {\bibinfo {author} {\bibfnamefont {E.}~\bibnamefont
  {Epelbaum}}, \bibinfo {author} {\bibfnamefont {H.}~\bibnamefont {Krebs}},
  \bibinfo {author} {\bibfnamefont {D.}~\bibnamefont {Lee}}, \ and\ \bibinfo
  {author} {\bibfnamefont {U.-G.}\ \bibnamefont {Meissner}},\ }\href {\doibase
  10.1103/PhysRevLett.106.192501} {\bibfield  {journal} {\bibinfo  {journal}
  {Phys. Rev. Lett.}\ }\textbf {\bibinfo {volume} {106}},\ \bibinfo {pages}
  {192501} (\bibinfo {year} {2011})},\ \Eprint {http://arxiv.org/abs/1101.2547}
  {arXiv:1101.2547 [nucl-th]} \BibitemShut {NoStop}%
\bibitem [{\citenamefont {Carlson}\ \emph {et~al.}(2015)\citenamefont
  {Carlson}, \citenamefont {Gandolfi}, \citenamefont {Pederiva}, \citenamefont
  {Pieper}, \citenamefont {Schiavilla}, \citenamefont {Schmidt},\ and\
  \citenamefont {Wiringa}}]{Carlson:2014vla}%
  \BibitemOpen
  \bibfield  {author} {\bibinfo {author} {\bibfnamefont {J.}~\bibnamefont
  {Carlson}}, \bibinfo {author} {\bibfnamefont {S.}~\bibnamefont {Gandolfi}},
  \bibinfo {author} {\bibfnamefont {F.}~\bibnamefont {Pederiva}}, \bibinfo
  {author} {\bibfnamefont {S.~C.}\ \bibnamefont {Pieper}}, \bibinfo {author}
  {\bibfnamefont {R.}~\bibnamefont {Schiavilla}}, \bibinfo {author}
  {\bibfnamefont {K.}~\bibnamefont {Schmidt}}, \ and\ \bibinfo {author}
  {\bibfnamefont {R.}~\bibnamefont {Wiringa}},\ }\href {\doibase
  10.1103/RevModPhys.87.1067} {\bibfield  {journal} {\bibinfo  {journal} {Rev.
  Mod. Phys.}\ }\textbf {\bibinfo {volume} {87}},\ \bibinfo {pages} {1067}
  (\bibinfo {year} {2015})},\ \Eprint {http://arxiv.org/abs/1412.3081}
  {arXiv:1412.3081 [nucl-th]} \BibitemShut {NoStop}%
\bibitem [{\citenamefont {Hu}\ \emph {et~al.}(2022)\citenamefont {Hu} \emph
  {et~al.}}]{Hu:2021trw}%
  \BibitemOpen
  \bibfield  {author} {\bibinfo {author} {\bibfnamefont {B.}~\bibnamefont {Hu}}
  \emph {et~al.},\ }\href {\doibase 10.1038/s41567-022-01715-8} {\bibfield
  {journal} {\bibinfo  {journal} {Nature Phys.}\ }\textbf {\bibinfo {volume}
  {18}},\ \bibinfo {pages} {1196} (\bibinfo {year} {2022})},\ \Eprint
  {http://arxiv.org/abs/2112.01125} {arXiv:2112.01125 [nucl-th]} \BibitemShut
  {NoStop}%
\bibitem [{\citenamefont {Gysbers}\ \emph {et~al.}(2019)\citenamefont {Gysbers}
  \emph {et~al.}}]{Gysbers:2019uyb}%
  \BibitemOpen
  \bibfield  {author} {\bibinfo {author} {\bibfnamefont {P.}~\bibnamefont
  {Gysbers}} \emph {et~al.},\ }\href {\doibase 10.1038/s41567-019-0450-7}
  {\bibfield  {journal} {\bibinfo  {journal} {Nature Phys.}\ }\textbf {\bibinfo
  {volume} {15}},\ \bibinfo {pages} {428} (\bibinfo {year} {2019})},\ \Eprint
  {http://arxiv.org/abs/1903.00047} {arXiv:1903.00047 [nucl-th]} \BibitemShut
  {NoStop}%
\bibitem [{\citenamefont {Roggero}\ and\ \citenamefont
  {Carlson}(2019)}]{Roggero:2018hrn}%
  \BibitemOpen
  \bibfield  {author} {\bibinfo {author} {\bibfnamefont {A.}~\bibnamefont
  {Roggero}}\ and\ \bibinfo {author} {\bibfnamefont {J.}~\bibnamefont
  {Carlson}},\ }\href {\doibase 10.1103/PhysRevC.100.034610} {\bibfield
  {journal} {\bibinfo  {journal} {Phys. Rev. C}\ }\textbf {\bibinfo {volume}
  {100}},\ \bibinfo {pages} {034610} (\bibinfo {year} {2019})},\ \Eprint
  {http://arxiv.org/abs/1804.01505} {arXiv:1804.01505 [quant-ph]} \BibitemShut
  {NoStop}%
\bibitem [{\citenamefont {Carleo}\ and\ \citenamefont
  {Troyer}(2017)}]{carleo:2017}%
  \BibitemOpen
  \bibfield  {author} {\bibinfo {author} {\bibfnamefont {G.}~\bibnamefont
  {Carleo}}\ and\ \bibinfo {author} {\bibfnamefont {M.}~\bibnamefont
  {Troyer}},\ }\href@noop {} {\bibfield  {journal} {\bibinfo  {journal}
  {Science}\ }\textbf {\bibinfo {volume} {355}},\ \bibinfo {pages} {602}
  (\bibinfo {year} {2017})}\BibitemShut {NoStop}%
\bibitem [{\citenamefont {{Schmitt}}\ and\ \citenamefont
  {{Heyl}}(2020)}]{Schmitt:2020}%
  \BibitemOpen
  \bibfield  {author} {\bibinfo {author} {\bibfnamefont {M.}~\bibnamefont
  {{Schmitt}}}\ and\ \bibinfo {author} {\bibfnamefont {M.}~\bibnamefont
  {{Heyl}}},\ }\href {\doibase 10.1103/PhysRevLett.125.100503} {\bibfield
  {journal} {\bibinfo  {journal} {\prl}\ }\textbf {\bibinfo {volume} {125}},\
  \bibinfo {eid} {100503} (\bibinfo {year} {2020})},\ \Eprint
  {http://arxiv.org/abs/1912.08828} {arXiv:1912.08828 [cond-mat.str-el]}
  \BibitemShut {NoStop}%
\bibitem [{\citenamefont {Miessen}\ \emph {et~al.}(2023)\citenamefont
  {Miessen}, \citenamefont {Ollitrault}, \citenamefont {Tacchino},\ and\
  \citenamefont {Tavernelli}}]{Miessen:2023}%
  \BibitemOpen
  \bibfield  {author} {\bibinfo {author} {\bibfnamefont {A.}~\bibnamefont
  {Miessen}}, \bibinfo {author} {\bibfnamefont {P.~J.}\ \bibnamefont
  {Ollitrault}}, \bibinfo {author} {\bibfnamefont {F.}~\bibnamefont
  {Tacchino}}, \ and\ \bibinfo {author} {\bibfnamefont {I.}~\bibnamefont
  {Tavernelli}},\ }\href {\doibase 10.1038/s43588-022-00374-2} {\bibfield
  {journal} {\bibinfo  {journal} {Nature Computational Science}\ }\textbf
  {\bibinfo {volume} {3}},\ \bibinfo {pages} {25} (\bibinfo {year}
  {2023})}\BibitemShut {NoStop}%
\bibitem [{\citenamefont {Hen}\ \emph {et~al.}(2017)\citenamefont {Hen},
  \citenamefont {Miller}, \citenamefont {Piasetzky},\ and\ \citenamefont
  {Weinstein}}]{Hen:2016kwk}%
  \BibitemOpen
  \bibfield  {author} {\bibinfo {author} {\bibfnamefont {O.}~\bibnamefont
  {Hen}}, \bibinfo {author} {\bibfnamefont {G.~A.}\ \bibnamefont {Miller}},
  \bibinfo {author} {\bibfnamefont {E.}~\bibnamefont {Piasetzky}}, \ and\
  \bibinfo {author} {\bibfnamefont {L.~B.}\ \bibnamefont {Weinstein}},\ }\href
  {\doibase 10.1103/RevModPhys.89.045002} {\bibfield  {journal} {\bibinfo
  {journal} {Rev. Mod. Phys.}\ }\textbf {\bibinfo {volume} {89}},\ \bibinfo
  {pages} {045002} (\bibinfo {year} {2017})},\ \Eprint
  {http://arxiv.org/abs/1611.09748} {arXiv:1611.09748 [nucl-ex]} \BibitemShut
  {NoStop}%
\bibitem [{\citenamefont {Segarra}\ \emph {et~al.}(2020)\citenamefont
  {Segarra}, \citenamefont {Schmidt}, \citenamefont {Kutz}, \citenamefont
  {Higinbotham}, \citenamefont {Piasetzky}, \citenamefont {Strikman},
  \citenamefont {Weinstein},\ and\ \citenamefont {Hen}}]{Segarra:2019gbp}%
  \BibitemOpen
  \bibfield  {author} {\bibinfo {author} {\bibfnamefont {E.~P.}\ \bibnamefont
  {Segarra}}, \bibinfo {author} {\bibfnamefont {A.}~\bibnamefont {Schmidt}},
  \bibinfo {author} {\bibfnamefont {T.}~\bibnamefont {Kutz}}, \bibinfo {author}
  {\bibfnamefont {D.~W.}\ \bibnamefont {Higinbotham}}, \bibinfo {author}
  {\bibfnamefont {E.}~\bibnamefont {Piasetzky}}, \bibinfo {author}
  {\bibfnamefont {M.}~\bibnamefont {Strikman}}, \bibinfo {author}
  {\bibfnamefont {L.~B.}\ \bibnamefont {Weinstein}}, \ and\ \bibinfo {author}
  {\bibfnamefont {O.}~\bibnamefont {Hen}},\ }\href {\doibase
  10.1103/PhysRevLett.124.092002} {\bibfield  {journal} {\bibinfo  {journal}
  {Phys. Rev. Lett.}\ }\textbf {\bibinfo {volume} {124}},\ \bibinfo {pages}
  {092002} (\bibinfo {year} {2020})},\ \Eprint
  {http://arxiv.org/abs/1908.02223} {arXiv:1908.02223 [nucl-th]} \BibitemShut
  {NoStop}%
\bibitem [{\citenamefont {Segarra}\ \emph {et~al.}(2021)\citenamefont
  {Segarra}, \citenamefont {Pybus}, \citenamefont {Hauenstein}, \citenamefont
  {Higinbotham}, \citenamefont {Miller}, \citenamefont {Piasetzky},
  \citenamefont {Schmidt}, \citenamefont {Strikman}, \citenamefont
  {Weinstein},\ and\ \citenamefont {Hen}}]{Segarra:2020plg}%
  \BibitemOpen
  \bibfield  {author} {\bibinfo {author} {\bibfnamefont {E.~P.}\ \bibnamefont
  {Segarra}}, \bibinfo {author} {\bibfnamefont {J.~R.}\ \bibnamefont {Pybus}},
  \bibinfo {author} {\bibfnamefont {F.}~\bibnamefont {Hauenstein}}, \bibinfo
  {author} {\bibfnamefont {D.~W.}\ \bibnamefont {Higinbotham}}, \bibinfo
  {author} {\bibfnamefont {G.~A.}\ \bibnamefont {Miller}}, \bibinfo {author}
  {\bibfnamefont {E.}~\bibnamefont {Piasetzky}}, \bibinfo {author}
  {\bibfnamefont {A.}~\bibnamefont {Schmidt}}, \bibinfo {author} {\bibfnamefont
  {M.}~\bibnamefont {Strikman}}, \bibinfo {author} {\bibfnamefont {L.~B.}\
  \bibnamefont {Weinstein}}, \ and\ \bibinfo {author} {\bibfnamefont
  {O.}~\bibnamefont {Hen}},\ }\href {\doibase 10.1103/PhysRevResearch.3.023240}
  {\bibfield  {journal} {\bibinfo  {journal} {Phys. Rev. Res.}\ }\textbf
  {\bibinfo {volume} {3}},\ \bibinfo {pages} {023240} (\bibinfo {year}
  {2021})},\ \Eprint {http://arxiv.org/abs/2006.10249} {arXiv:2006.10249
  [hep-ph]} \BibitemShut {NoStop}%
\bibitem [{\citenamefont {Benhar}\ \emph {et~al.}(2017)\citenamefont {Benhar},
  \citenamefont {Huber}, \citenamefont {Mariani},\ and\ \citenamefont
  {Meloni}}]{Benhar:2015wva}%
  \BibitemOpen
  \bibfield  {author} {\bibinfo {author} {\bibfnamefont {O.}~\bibnamefont
  {Benhar}}, \bibinfo {author} {\bibfnamefont {P.}~\bibnamefont {Huber}},
  \bibinfo {author} {\bibfnamefont {C.}~\bibnamefont {Mariani}}, \ and\
  \bibinfo {author} {\bibfnamefont {D.}~\bibnamefont {Meloni}},\ }\href
  {\doibase 10.1016/j.physrep.2017.07.004} {\bibfield  {journal} {\bibinfo
  {journal} {Phys. Rept.}\ }\textbf {\bibinfo {volume} {700}},\ \bibinfo
  {pages} {1} (\bibinfo {year} {2017})},\ \Eprint
  {http://arxiv.org/abs/1501.06448} {arXiv:1501.06448 [nucl-th]} \BibitemShut
  {NoStop}%
%%CITATION = ARXIV:1501.06448;%%
\bibitem [{\citenamefont {Katori}\ and\ \citenamefont
  {Martini}(2018)}]{Katori:2016yel}%
  \BibitemOpen
  \bibfield  {author} {\bibinfo {author} {\bibfnamefont {T.}~\bibnamefont
  {Katori}}\ and\ \bibinfo {author} {\bibfnamefont {M.}~\bibnamefont
  {Martini}},\ }\href {\doibase 10.1088/1361-6471/aa8bf7} {\bibfield  {journal}
  {\bibinfo  {journal} {J. Phys. G}\ }\textbf {\bibinfo {volume} {45}},\
  \bibinfo {pages} {013001} (\bibinfo {year} {2018})},\ \Eprint
  {http://arxiv.org/abs/1611.07770} {arXiv:1611.07770 [hep-ph]} \BibitemShut
  {NoStop}%
%%CITATION = ARXIV:1611.07770;%%
\bibitem [{\citenamefont {Alvarez-Ruso}\ \emph {et~al.}(2018)\citenamefont
  {Alvarez-Ruso} \emph {et~al.}}]{NuSTEC:2017hzk}%
  \BibitemOpen
  \bibfield  {author} {\bibinfo {author} {\bibfnamefont {L.}~\bibnamefont
  {Alvarez-Ruso}} \emph {et~al.} (\bibinfo {collaboration} {NuSTEC}),\ }\href
  {\doibase 10.1016/j.ppnp.2018.01.006} {\bibfield  {journal} {\bibinfo
  {journal} {Prog. Part. Nucl. Phys.}\ }\textbf {\bibinfo {volume} {100}},\
  \bibinfo {pages} {1} (\bibinfo {year} {2018})},\ \Eprint
  {http://arxiv.org/abs/1706.03621} {arXiv:1706.03621 [hep-ph]} \BibitemShut
  {NoStop}%
\bibitem [{\citenamefont {Lovato}\ \emph {et~al.}(2016)\citenamefont {Lovato},
  \citenamefont {Gandolfi}, \citenamefont {Carlson}, \citenamefont {Pieper},\
  and\ \citenamefont {Schiavilla}}]{Lovato:2016gkq}%
  \BibitemOpen
  \bibfield  {author} {\bibinfo {author} {\bibfnamefont {A.}~\bibnamefont
  {Lovato}}, \bibinfo {author} {\bibfnamefont {S.}~\bibnamefont {Gandolfi}},
  \bibinfo {author} {\bibfnamefont {J.}~\bibnamefont {Carlson}}, \bibinfo
  {author} {\bibfnamefont {S.~C.}\ \bibnamefont {Pieper}}, \ and\ \bibinfo
  {author} {\bibfnamefont {R.}~\bibnamefont {Schiavilla}},\ }\href {\doibase
  10.1103/PhysRevLett.117.082501} {\bibfield  {journal} {\bibinfo  {journal}
  {Phys. Rev. Lett.}\ }\textbf {\bibinfo {volume} {117}},\ \bibinfo {pages}
  {082501} (\bibinfo {year} {2016})},\ \Eprint
  {http://arxiv.org/abs/1605.00248} {arXiv:1605.00248 [nucl-th]} \BibitemShut
  {NoStop}%
\bibitem [{\citenamefont {Lovato}\ \emph {et~al.}(2020)\citenamefont {Lovato},
  \citenamefont {Carlson}, \citenamefont {Gandolfi}, \citenamefont {Rocco},\
  and\ \citenamefont {Schiavilla}}]{Lovato:2020kba}%
  \BibitemOpen
  \bibfield  {author} {\bibinfo {author} {\bibfnamefont {A.}~\bibnamefont
  {Lovato}}, \bibinfo {author} {\bibfnamefont {J.}~\bibnamefont {Carlson}},
  \bibinfo {author} {\bibfnamefont {S.}~\bibnamefont {Gandolfi}}, \bibinfo
  {author} {\bibfnamefont {N.}~\bibnamefont {Rocco}}, \ and\ \bibinfo {author}
  {\bibfnamefont {R.}~\bibnamefont {Schiavilla}},\ }\href {\doibase
  10.1103/PhysRevX.10.031068} {\bibfield  {journal} {\bibinfo  {journal} {Phys.
  Rev. X}\ }\textbf {\bibinfo {volume} {10}},\ \bibinfo {pages} {031068}
  (\bibinfo {year} {2020})},\ \Eprint {http://arxiv.org/abs/2003.07710}
  {arXiv:2003.07710 [nucl-th]} \BibitemShut {NoStop}%
\bibitem [{\citenamefont {Sobczyk}\ \emph {et~al.}(2021)\citenamefont
  {Sobczyk}, \citenamefont {Acharya}, \citenamefont {Bacca},\ and\
  \citenamefont {Hagen}}]{Sobczyk:2021dwm}%
  \BibitemOpen
  \bibfield  {author} {\bibinfo {author} {\bibfnamefont {J.~E.}\ \bibnamefont
  {Sobczyk}}, \bibinfo {author} {\bibfnamefont {B.}~\bibnamefont {Acharya}},
  \bibinfo {author} {\bibfnamefont {S.}~\bibnamefont {Bacca}}, \ and\ \bibinfo
  {author} {\bibfnamefont {G.}~\bibnamefont {Hagen}},\ }\href {\doibase
  10.1103/PhysRevLett.127.072501} {\bibfield  {journal} {\bibinfo  {journal}
  {Phys. Rev. Lett.}\ }\textbf {\bibinfo {volume} {127}},\ \bibinfo {pages}
  {072501} (\bibinfo {year} {2021})},\ \Eprint
  {http://arxiv.org/abs/2103.06786} {arXiv:2103.06786 [nucl-th]} \BibitemShut
  {NoStop}%
\bibitem [{\citenamefont {Sobczyk}\ and\ \citenamefont
  {Roggero}(2022)}]{Sobczyk:2021ejs}%
  \BibitemOpen
  \bibfield  {author} {\bibinfo {author} {\bibfnamefont {J.~E.}\ \bibnamefont
  {Sobczyk}}\ and\ \bibinfo {author} {\bibfnamefont {A.}~\bibnamefont
  {Roggero}},\ }\href {\doibase 10.1103/PhysRevE.105.055310} {\bibfield
  {journal} {\bibinfo  {journal} {Phys. Rev. E}\ }\textbf {\bibinfo {volume}
  {105}},\ \bibinfo {pages} {055310} (\bibinfo {year} {2022})},\ \Eprint
  {http://arxiv.org/abs/2110.02108} {arXiv:2110.02108 [nucl-th]} \BibitemShut
  {NoStop}%
\bibitem [{\citenamefont {Carlson}\ \emph {et~al.}(2002)\citenamefont
  {Carlson}, \citenamefont {Jourdan}, \citenamefont {Schiavilla},\ and\
  \citenamefont {Sick}}]{Carlson:2001mp}%
  \BibitemOpen
  \bibfield  {author} {\bibinfo {author} {\bibfnamefont {J.}~\bibnamefont
  {Carlson}}, \bibinfo {author} {\bibfnamefont {J.}~\bibnamefont {Jourdan}},
  \bibinfo {author} {\bibfnamefont {R.}~\bibnamefont {Schiavilla}}, \ and\
  \bibinfo {author} {\bibfnamefont {I.}~\bibnamefont {Sick}},\ }\href {\doibase
  10.1103/PhysRevC.65.024002} {\bibfield  {journal} {\bibinfo  {journal} {Phys.
  Rev. C}\ }\textbf {\bibinfo {volume} {65}},\ \bibinfo {pages} {024002}
  (\bibinfo {year} {2002})},\ \Eprint {http://arxiv.org/abs/nucl-th/0106047}
  {arXiv:nucl-th/0106047} \BibitemShut {NoStop}%
\bibitem [{\citenamefont {Bryan}(1990)}]{Bryan:1990}%
  \BibitemOpen
  \bibfield  {author} {\bibinfo {author} {\bibfnamefont {R.}~\bibnamefont
  {Bryan}},\ }\href {\doibase 10.1007/BF02427376} {\bibfield  {journal}
  {\bibinfo  {journal} {European Biophysics Journal}\ }\textbf {\bibinfo
  {volume} {18}},\ \bibinfo {pages} {165} (\bibinfo {year} {1990})}\BibitemShut
  {NoStop}%
\bibitem [{\citenamefont {Jarrell}\ and\ \citenamefont
  {Gubernatis}(1996)}]{Jarrell:1996}%
  \BibitemOpen
  \bibfield  {author} {\bibinfo {author} {\bibfnamefont {M.}~\bibnamefont
  {Jarrell}}\ and\ \bibinfo {author} {\bibfnamefont {J.}~\bibnamefont
  {Gubernatis}},\ }\href {\doibase
  http://dx.doi.org/10.1016/0370-1573(95)00074-7} {\bibfield  {journal}
  {\bibinfo  {journal} {Physics Reports}\ }\textbf {\bibinfo {volume} {269}},\
  \bibinfo {pages} {133 } (\bibinfo {year} {1996})}\BibitemShut {NoStop}%
\bibitem [{\citenamefont {Lovato}\ \emph {et~al.}(2019)\citenamefont {Lovato},
  \citenamefont {Rocco},\ and\ \citenamefont {Schiavilla}}]{Lovato:2019fiw}%
  \BibitemOpen
  \bibfield  {author} {\bibinfo {author} {\bibfnamefont {A.}~\bibnamefont
  {Lovato}}, \bibinfo {author} {\bibfnamefont {N.}~\bibnamefont {Rocco}}, \
  and\ \bibinfo {author} {\bibfnamefont {R.}~\bibnamefont {Schiavilla}},\
  }\href {\doibase 10.1103/PhysRevC.100.035502} {\bibfield  {journal} {\bibinfo
   {journal} {Phys. Rev. C}\ }\textbf {\bibinfo {volume} {100}},\ \bibinfo
  {pages} {035502} (\bibinfo {year} {2019})},\ \Eprint
  {http://arxiv.org/abs/1903.08078} {arXiv:1903.08078 [nucl-th]} \BibitemShut
  {NoStop}%
\bibitem [{\citenamefont {{McCann}}\ \emph {et~al.}(2017)\citenamefont
  {{McCann}}, \citenamefont {{Jin}},\ and\ \citenamefont
  {{Unser}}}]{McCann:2017}%
  \BibitemOpen
  \bibfield  {author} {\bibinfo {author} {\bibfnamefont {M.~T.}\ \bibnamefont
  {{McCann}}}, \bibinfo {author} {\bibfnamefont {K.~H.}\ \bibnamefont {{Jin}}},
  \ and\ \bibinfo {author} {\bibfnamefont {M.}~\bibnamefont {{Unser}}},\ }\href
  {\doibase 10.1109/MSP.2017.2739299} {\bibfield  {journal} {\bibinfo
  {journal} {IEEE Signal Process. Mag.}\ }\textbf {\bibinfo {volume} {34}},\
  \bibinfo {pages} {85} (\bibinfo {year} {2017})},\ \Eprint
  {http://arxiv.org/abs/1710.04011} {arXiv:1710.04011 [eess.IV]} \BibitemShut
  {NoStop}%
\bibitem [{\citenamefont {{Arsenault}}\ \emph {et~al.}(2017)\citenamefont
  {{Arsenault}}, \citenamefont {{Neuberg}}, \citenamefont {{Hannah}},\ and\
  \citenamefont {{Millis}}}]{Arsenault:2017}%
  \BibitemOpen
  \bibfield  {author} {\bibinfo {author} {\bibfnamefont {L.-F.}\ \bibnamefont
  {{Arsenault}}}, \bibinfo {author} {\bibfnamefont {R.}~\bibnamefont
  {{Neuberg}}}, \bibinfo {author} {\bibfnamefont {L.~A.}\ \bibnamefont
  {{Hannah}}}, \ and\ \bibinfo {author} {\bibfnamefont {A.~J.}\ \bibnamefont
  {{Millis}}},\ }\href {\doibase 10.1088/1361-6420/aa8d93} {\bibfield
  {journal} {\bibinfo  {journal} {Inverse Probl.}\ }\textbf {\bibinfo {volume}
  {33}},\ \bibinfo {eid} {115007} (\bibinfo {year} {2017})}\BibitemShut
  {NoStop}%
\bibitem [{\citenamefont {{Yoon}}\ \emph {et~al.}(2018)\citenamefont {{Yoon}},
  \citenamefont {{Sim}},\ and\ \citenamefont {{Han}}}]{Yoon:2018}%
  \BibitemOpen
  \bibfield  {author} {\bibinfo {author} {\bibfnamefont {H.}~\bibnamefont
  {{Yoon}}}, \bibinfo {author} {\bibfnamefont {J.-H.}\ \bibnamefont {{Sim}}}, \
  and\ \bibinfo {author} {\bibfnamefont {M.~J.}\ \bibnamefont {{Han}}},\ }\href
  {\doibase 10.1103/PhysRevB.98.245101} {\bibfield  {journal} {\bibinfo
  {journal} {\prb}\ }\textbf {\bibinfo {volume} {98}},\ \bibinfo {eid} {245101}
  (\bibinfo {year} {2018})},\ \Eprint {http://arxiv.org/abs/1806.03841}
  {arXiv:1806.03841 [cond-mat.str-el]} \BibitemShut {NoStop}%
\bibitem [{\citenamefont {{Xie}}\ \emph {et~al.}(2019)\citenamefont {{Xie}},
  \citenamefont {{Bao}}, \citenamefont {{Maier}},\ and\ \citenamefont
  {{Webster}}}]{Xie:2019}%
  \BibitemOpen
  \bibfield  {author} {\bibinfo {author} {\bibfnamefont {X.}~\bibnamefont
  {{Xie}}}, \bibinfo {author} {\bibfnamefont {F.}~\bibnamefont {{Bao}}},
  \bibinfo {author} {\bibfnamefont {T.}~\bibnamefont {{Maier}}}, \ and\
  \bibinfo {author} {\bibfnamefont {C.}~\bibnamefont {{Webster}}},\ }\href@noop
  {} {\bibfield  {journal} {\bibinfo  {journal} {arXiv}\ } (\bibinfo {year}
  {2019})},\ \Eprint {http://arxiv.org/abs/1905.10430} {arXiv:1905.10430
  [physics.comp-ph]} \BibitemShut {NoStop}%
\bibitem [{\citenamefont {{Fournier}}\ \emph {et~al.}(2020)\citenamefont
  {{Fournier}}, \citenamefont {{Wang}}, \citenamefont {{Yazyev}},\ and\
  \citenamefont {{Wu}}}]{Fournier:2020}%
  \BibitemOpen
  \bibfield  {author} {\bibinfo {author} {\bibfnamefont {R.}~\bibnamefont
  {{Fournier}}}, \bibinfo {author} {\bibfnamefont {L.}~\bibnamefont {{Wang}}},
  \bibinfo {author} {\bibfnamefont {O.~V.}\ \bibnamefont {{Yazyev}}}, \ and\
  \bibinfo {author} {\bibfnamefont {Q.}~\bibnamefont {{Wu}}},\ }\href {\doibase
  10.1103/PhysRevLett.124.056401} {\bibfield  {journal} {\bibinfo  {journal}
  {\prl}\ }\textbf {\bibinfo {volume} {124}},\ \bibinfo {eid} {056401}
  (\bibinfo {year} {2020})}\BibitemShut {NoStop}%
\bibitem [{\citenamefont {Raghavan}\ \emph {et~al.}(2021)\citenamefont
  {Raghavan}, \citenamefont {Balaprakash}, \citenamefont {Lovato},
  \citenamefont {Rocco},\ and\ \citenamefont {Wild}}]{Raghavan:2020bze}%
  \BibitemOpen
  \bibfield  {author} {\bibinfo {author} {\bibfnamefont {K.}~\bibnamefont
  {Raghavan}}, \bibinfo {author} {\bibfnamefont {P.}~\bibnamefont
  {Balaprakash}}, \bibinfo {author} {\bibfnamefont {A.}~\bibnamefont {Lovato}},
  \bibinfo {author} {\bibfnamefont {N.}~\bibnamefont {Rocco}}, \ and\ \bibinfo
  {author} {\bibfnamefont {S.~M.}\ \bibnamefont {Wild}},\ }\href {\doibase
  10.1103/PhysRevC.103.035502} {\bibfield  {journal} {\bibinfo  {journal}
  {Phys. Rev. C}\ }\textbf {\bibinfo {volume} {103}},\ \bibinfo {pages}
  {035502} (\bibinfo {year} {2021})},\ \Eprint
  {http://arxiv.org/abs/2010.12703} {arXiv:2010.12703 [nucl-th]} \BibitemShut
  {NoStop}%
\bibitem [{\citenamefont {Shen}\ \emph {et~al.}(2012)\citenamefont {Shen},
  \citenamefont {Marcucci}, \citenamefont {Carlson}, \citenamefont {Gandolfi},\
  and\ \citenamefont {Schiavilla}}]{Shen:2012xz}%
  \BibitemOpen
  \bibfield  {author} {\bibinfo {author} {\bibfnamefont {G.}~\bibnamefont
  {Shen}}, \bibinfo {author} {\bibfnamefont {L.~E.}\ \bibnamefont {Marcucci}},
  \bibinfo {author} {\bibfnamefont {J.}~\bibnamefont {Carlson}}, \bibinfo
  {author} {\bibfnamefont {S.}~\bibnamefont {Gandolfi}}, \ and\ \bibinfo
  {author} {\bibfnamefont {R.}~\bibnamefont {Schiavilla}},\ }\href {\doibase
  10.1103/PhysRevC.86.035503} {\bibfield  {journal} {\bibinfo  {journal} {Phys.
  Rev. C}\ }\textbf {\bibinfo {volume} {86}},\ \bibinfo {pages} {035503}
  (\bibinfo {year} {2012})},\ \Eprint {http://arxiv.org/abs/1205.4337}
  {arXiv:1205.4337 [nucl-th]} \BibitemShut {NoStop}%
%%CITATION = ARXIV:1205.4337;%%
\bibitem [{\citenamefont {Golak}\ \emph {et~al.}(2018)\citenamefont {Golak},
  \citenamefont {Skibiński}, \citenamefont {Topolnicki}, \citenamefont
  {Witała}, \citenamefont {Grassi}, \citenamefont {Kamada},\ and\
  \citenamefont {Marcucci}}]{Golak:2018qya}%
  \BibitemOpen
  \bibfield  {author} {\bibinfo {author} {\bibfnamefont {J.}~\bibnamefont
  {Golak}}, \bibinfo {author} {\bibfnamefont {R.}~\bibnamefont {Skibiński}},
  \bibinfo {author} {\bibfnamefont {K.}~\bibnamefont {Topolnicki}}, \bibinfo
  {author} {\bibfnamefont {H.}~\bibnamefont {Witała}}, \bibinfo {author}
  {\bibfnamefont {A.}~\bibnamefont {Grassi}}, \bibinfo {author} {\bibfnamefont
  {H.}~\bibnamefont {Kamada}}, \ and\ \bibinfo {author} {\bibfnamefont {L.~E.}\
  \bibnamefont {Marcucci}},\ }\href {\doibase 10.1103/PhysRevC.98.015501}
  {\bibfield  {journal} {\bibinfo  {journal} {Phys. Rev. C}\ }\textbf {\bibinfo
  {volume} {C8}},\ \bibinfo {pages} {015501} (\bibinfo {year} {2018})},\
  \Eprint {http://arxiv.org/abs/1805.00103} {arXiv:1805.00103 [nucl-th]}
  \BibitemShut {NoStop}%
%%CITATION = ARXIV:1805.00103;%%
\bibitem [{\citenamefont {Carlson}\ and\ \citenamefont
  {Schiavilla}(1992)}]{Carlson:1992ga}%
  \BibitemOpen
  \bibfield  {author} {\bibinfo {author} {\bibfnamefont {J.}~\bibnamefont
  {Carlson}}\ and\ \bibinfo {author} {\bibfnamefont {R.}~\bibnamefont
  {Schiavilla}},\ }\href {\doibase 10.1103/PhysRevLett.68.3682} {\bibfield
  {journal} {\bibinfo  {journal} {Phys. Rev. Lett.}\ }\textbf {\bibinfo
  {volume} {68}},\ \bibinfo {pages} {3682} (\bibinfo {year}
  {1992})}\BibitemShut {NoStop}%
%%CITATION = PRLTA,68,3682;%%
\bibitem [{\citenamefont {{Gull}}\ and\ \citenamefont
  {{Daniell}}(1978)}]{Gull:1978}%
  \BibitemOpen
  \bibfield  {author} {\bibinfo {author} {\bibfnamefont {S.~F.}\ \bibnamefont
  {{Gull}}}\ and\ \bibinfo {author} {\bibfnamefont {G.~J.}\ \bibnamefont
  {{Daniell}}},\ }\href {\doibase 10.1038/272686a0} {\bibfield  {journal}
  {\bibinfo  {journal} {\nat}\ }\textbf {\bibinfo {volume} {272}},\ \bibinfo
  {pages} {686} (\bibinfo {year} {1978})}\BibitemShut {NoStop}%
\bibitem [{\citenamefont {Skilling}(1989)}]{Skilling:1989}%
  \BibitemOpen
  \bibfield  {author} {\bibinfo {author} {\bibfnamefont {J.}~\bibnamefont
  {Skilling}},\ }\enquote {\bibinfo {title} {Classic maximum entropy},}\ in\
  \href {\doibase 10.1007/978-94-015-7860-8_3} {\emph {\bibinfo {booktitle}
  {Maximum Entropy and Bayesian Method}}},\ \bibinfo {editor} {edited by\
  \bibinfo {editor} {\bibfnamefont {J.}~\bibnamefont {Skilling}}}\ (\bibinfo
  {publisher} {Springer Netherlands},\ \bibinfo {address} {Dordrecht},\
  \bibinfo {year} {1989})\ pp.\ \bibinfo {pages} {45--52}\BibitemShut {NoStop}%
\bibitem [{\citenamefont {{Von Der Linden}}\ \emph {et~al.}(1999)\citenamefont
  {{Von Der Linden}}, \citenamefont {Preuss},\ and\ \citenamefont
  {Dose}}]{VonDerLinden:1999}%
  \BibitemOpen
  \bibfield  {author} {\bibinfo {author} {\bibfnamefont {W.}~\bibnamefont {{Von
  Der Linden}}}, \bibinfo {author} {\bibfnamefont {R.}~\bibnamefont {Preuss}},
  \ and\ \bibinfo {author} {\bibfnamefont {V.}~\bibnamefont {Dose}},\ }in\
  \href@noop {} {\emph {\bibinfo {booktitle} {Maximum Entropy and Bayesian
  Methods}}},\ \bibinfo {editor} {edited by\ \bibinfo {editor} {\bibfnamefont
  {W.}~\bibnamefont {von~der Linden}}, \bibinfo {editor} {\bibfnamefont
  {V.}~\bibnamefont {Dose}}, \bibinfo {editor} {\bibfnamefont {R.}~\bibnamefont
  {Fischer}}, \ and\ \bibinfo {editor} {\bibfnamefont {R.}~\bibnamefont
  {Preuss}}}\ (\bibinfo  {publisher} {Springer Netherlands},\ \bibinfo
  {address} {Dordrecht},\ \bibinfo {year} {1999})\ pp.\ \bibinfo {pages}
  {319--326}\BibitemShut {NoStop}%
\bibitem [{\citenamefont {{Hohenadler}}\ \emph {et~al.}(2005)\citenamefont
  {{Hohenadler}}, \citenamefont {{Neuber}}, \citenamefont {{von der Linden}},
  \citenamefont {{Wellein}}, \citenamefont {{Loos}},\ and\ \citenamefont
  {{Fehske}}}]{Hohenadler:2005}%
  \BibitemOpen
  \bibfield  {author} {\bibinfo {author} {\bibfnamefont {M.}~\bibnamefont
  {{Hohenadler}}}, \bibinfo {author} {\bibfnamefont {D.}~\bibnamefont
  {{Neuber}}}, \bibinfo {author} {\bibfnamefont {W.}~\bibnamefont {{von der
  Linden}}}, \bibinfo {author} {\bibfnamefont {G.}~\bibnamefont {{Wellein}}},
  \bibinfo {author} {\bibfnamefont {J.}~\bibnamefont {{Loos}}}, \ and\ \bibinfo
  {author} {\bibfnamefont {H.}~\bibnamefont {{Fehske}}},\ }\href {\doibase
  10.1103/PhysRevB.71.245111} {\bibfield  {journal} {\bibinfo  {journal}
  {\prb}\ }\textbf {\bibinfo {volume} {71}},\ \bibinfo {eid} {245111} (\bibinfo
  {year} {2005})},\ \Eprint {http://arxiv.org/abs/cond-mat/0412010}
  {arXiv:cond-mat/0412010 [cond-mat.str-el]} \BibitemShut {NoStop}%
\bibitem [{\citenamefont {Kullback}\ and\ \citenamefont
  {Leibler}(1951)}]{kullback1951information}%
  \BibitemOpen
  \bibfield  {author} {\bibinfo {author} {\bibfnamefont {S.}~\bibnamefont
  {Kullback}}\ and\ \bibinfo {author} {\bibfnamefont {R.~A.}\ \bibnamefont
  {Leibler}},\ }\href@noop {} {\bibfield  {journal} {\bibinfo  {journal} {The
  annals of mathematical statistics}\ }\textbf {\bibinfo {volume} {22}},\
  \bibinfo {pages} {79} (\bibinfo {year} {1951})}\BibitemShut {NoStop}%
\bibitem [{\citenamefont {Kingma}\ and\ \citenamefont
  {Ba}(2014)}]{kingma2014adam}%
  \BibitemOpen
  \bibfield  {author} {\bibinfo {author} {\bibfnamefont {D.~P.}\ \bibnamefont
  {Kingma}}\ and\ \bibinfo {author} {\bibfnamefont {J.}~\bibnamefont {Ba}},\
  }\href@noop {} {\bibfield  {journal} {\bibinfo  {journal} {arXiv preprint
  arXiv:1412.6980}\ } (\bibinfo {year} {2014})}\BibitemShut {NoStop}%
\bibitem [{\citenamefont {Ruso}\ \emph {et~al.}(2022)\citenamefont {Ruso} \emph
  {et~al.}}]{Ruso:2022qes}%
  \BibitemOpen
  \bibfield  {author} {\bibinfo {author} {\bibfnamefont {L.~A.}\ \bibnamefont
  {Ruso}} \emph {et~al.},\ }\href@noop {} {\enquote {\bibinfo {title}
  {{Theoretical tools for neutrino scattering: interplay between lattice QCD,
  EFTs, nuclear physics, phenomenology, and neutrino event generators}},}\ }
  (\bibinfo {year} {2022}),\ \bibinfo {note} {contribution to: Snowmass 2021},\
  \Eprint {http://arxiv.org/abs/2203.09030} {arXiv:2203.09030 [hep-ph]}
  \BibitemShut {NoStop}%
\bibitem [{\citenamefont {Zhang}(2021)}]{Zhang:2021}%
  \BibitemOpen
  \bibfield  {author} {\bibinfo {author} {\bibfnamefont {J.}~\bibnamefont
  {Zhang}},\ }\href {\doibase https://doi.org/10.1002/wics.1539} {\bibfield
  {journal} {\bibinfo  {journal} {WIREs Computational Statistics}\ }\textbf
  {\bibinfo {volume} {13}},\ \bibinfo {pages} {e1539} (\bibinfo {year}
  {2021})},\ \Eprint
  {http://arxiv.org/abs/https://wires.onlinelibrary.wiley.com/doi/pdf/10.1002/wics.1539}
  {https://wires.onlinelibrary.wiley.com/doi/pdf/10.1002/wics.1539}
  \BibitemShut {NoStop}%
\bibitem [{\citenamefont {Schmidt}\ and\ \citenamefont
  {Fantoni}(1999)}]{Schmidt:1999lik}%
  \BibitemOpen
  \bibfield  {author} {\bibinfo {author} {\bibfnamefont {K.}~\bibnamefont
  {Schmidt}}\ and\ \bibinfo {author} {\bibfnamefont {S.}~\bibnamefont
  {Fantoni}},\ }\href {\doibase 10.1016/S0370-2693(98)01522-6} {\bibfield
  {journal} {\bibinfo  {journal} {Phys. Lett. B}\ }\textbf {\bibinfo {volume}
  {446}},\ \bibinfo {pages} {99} (\bibinfo {year} {1999})}\BibitemShut
  {NoStop}%
\bibitem [{\citenamefont {Utama}\ \emph {et~al.}(2016)\citenamefont {Utama},
  \citenamefont {Piekarewicz},\ and\ \citenamefont {Prosper}}]{Utama:2015hva}%
  \BibitemOpen
  \bibfield  {author} {\bibinfo {author} {\bibfnamefont {R.}~\bibnamefont
  {Utama}}, \bibinfo {author} {\bibfnamefont {J.}~\bibnamefont {Piekarewicz}},
  \ and\ \bibinfo {author} {\bibfnamefont {H.}~\bibnamefont {Prosper}},\ }\href
  {\doibase 10.1103/PhysRevC.93.014311} {\bibfield  {journal} {\bibinfo
  {journal} {Phys. Rev. C}\ }\textbf {\bibinfo {volume} {93}},\ \bibinfo
  {pages} {014311} (\bibinfo {year} {2016})},\ \Eprint
  {http://arxiv.org/abs/1508.06263} {arXiv:1508.06263 [nucl-th]} \BibitemShut
  {NoStop}%
\bibitem [{\citenamefont {Niu}\ and\ \citenamefont
  {Liang}(2018)}]{Niu:2018csp}%
  \BibitemOpen
  \bibfield  {author} {\bibinfo {author} {\bibfnamefont {Z.~M.}\ \bibnamefont
  {Niu}}\ and\ \bibinfo {author} {\bibfnamefont {H.~Z.}\ \bibnamefont
  {Liang}},\ }\href {\doibase 10.1016/j.physletb.2018.01.002} {\bibfield
  {journal} {\bibinfo  {journal} {Phys. Lett. B}\ }\textbf {\bibinfo {volume}
  {778}},\ \bibinfo {pages} {48} (\bibinfo {year} {2018})},\ \Eprint
  {http://arxiv.org/abs/1801.04411} {arXiv:1801.04411 [nucl-th]} \BibitemShut
  {NoStop}%
\bibitem [{\citenamefont {Donnelly}\ and\ \citenamefont
  {Walecka}(1975)}]{Donnelly:1975ze}%
  \BibitemOpen
  \bibfield  {author} {\bibinfo {author} {\bibfnamefont {T.~W.}\ \bibnamefont
  {Donnelly}}\ and\ \bibinfo {author} {\bibfnamefont {J.~D.}\ \bibnamefont
  {Walecka}},\ }\href {\doibase 10.1146/annurev.ns.25.120175.001553} {\bibfield
   {journal} {\bibinfo  {journal} {Ann. Rev. Nucl. Part. Sci.}\ }\textbf
  {\bibinfo {volume} {25}},\ \bibinfo {pages} {329} (\bibinfo {year}
  {1975})}\BibitemShut {NoStop}%
\bibitem [{\citenamefont {Benhar}\ \emph {et~al.}(2008)\citenamefont {Benhar},
  \citenamefont {Day},\ and\ \citenamefont {Sick}}]{Benhar:2006wy}%
  \BibitemOpen
  \bibfield  {author} {\bibinfo {author} {\bibfnamefont {O.}~\bibnamefont
  {Benhar}}, \bibinfo {author} {\bibfnamefont {D.}~\bibnamefont {Day}}, \ and\
  \bibinfo {author} {\bibfnamefont {I.}~\bibnamefont {Sick}},\ }\href {\doibase
  10.1103/RevModPhys.80.189} {\bibfield  {journal} {\bibinfo  {journal} {Rev.\
  Mod.\ Phys.}\ }\textbf {\bibinfo {volume} {80}},\ \bibinfo {pages} {189}
  (\bibinfo {year} {2008})},\ \Eprint {http://arxiv.org/abs/nucl-ex/0603029}
  {arXiv:nucl-ex/0603029} \BibitemShut {NoStop}%
\bibitem [{\citenamefont {Kegel}\ \emph {et~al.}(2023)\citenamefont {Kegel}
  \emph {et~al.}}]{Kegel:2021jrh}%
  \BibitemOpen
  \bibfield  {author} {\bibinfo {author} {\bibfnamefont {S.}~\bibnamefont
  {Kegel}} \emph {et~al.},\ }\href {\doibase 10.1103/PhysRevLett.130.152502}
  {\bibfield  {journal} {\bibinfo  {journal} {Phys. Rev. Lett.}\ }\textbf
  {\bibinfo {volume} {130}},\ \bibinfo {pages} {152502} (\bibinfo {year}
  {2023})},\ \Eprint {http://arxiv.org/abs/2112.10582} {arXiv:2112.10582
  [nucl-ex]} \BibitemShut {NoStop}%
\bibitem [{\citenamefont {Adhikari}\ \emph {et~al.}(2021)\citenamefont
  {Adhikari} \emph {et~al.}}]{PREX:2021umo}%
  \BibitemOpen
  \bibfield  {author} {\bibinfo {author} {\bibfnamefont {D.}~\bibnamefont
  {Adhikari}} \emph {et~al.} (\bibinfo {collaboration} {PREX}),\ }\href
  {\doibase 10.1103/PhysRevLett.126.172502} {\bibfield  {journal} {\bibinfo
  {journal} {Phys. Rev. Lett.}\ }\textbf {\bibinfo {volume} {126}},\ \bibinfo
  {pages} {172502} (\bibinfo {year} {2021})},\ \Eprint
  {http://arxiv.org/abs/2102.10767} {arXiv:2102.10767 [nucl-ex]} \BibitemShut
  {NoStop}%
\bibitem [{\citenamefont {Adhikari}\ \emph {et~al.}(2022)\citenamefont
  {Adhikari} \emph {et~al.}}]{CREX:2022kgg}%
  \BibitemOpen
  \bibfield  {author} {\bibinfo {author} {\bibfnamefont {D.}~\bibnamefont
  {Adhikari}} \emph {et~al.} (\bibinfo {collaboration} {CREX}),\ }\href
  {\doibase 10.1103/PhysRevLett.129.042501} {\bibfield  {journal} {\bibinfo
  {journal} {Phys. Rev. Lett.}\ }\textbf {\bibinfo {volume} {129}},\ \bibinfo
  {pages} {042501} (\bibinfo {year} {2022})},\ \Eprint
  {http://arxiv.org/abs/2205.11593} {arXiv:2205.11593 [nucl-ex]} \BibitemShut
  {NoStop}%
\bibitem [{\citenamefont {Andreoli}\ \emph {et~al.}(2022)\citenamefont
  {Andreoli}, \citenamefont {Carlson}, \citenamefont {Lovato}, \citenamefont
  {Pastore}, \citenamefont {Rocco},\ and\ \citenamefont
  {Wiringa}}]{Andreoli:2021cxo}%
  \BibitemOpen
  \bibfield  {author} {\bibinfo {author} {\bibfnamefont {L.}~\bibnamefont
  {Andreoli}}, \bibinfo {author} {\bibfnamefont {J.}~\bibnamefont {Carlson}},
  \bibinfo {author} {\bibfnamefont {A.}~\bibnamefont {Lovato}}, \bibinfo
  {author} {\bibfnamefont {S.}~\bibnamefont {Pastore}}, \bibinfo {author}
  {\bibfnamefont {N.}~\bibnamefont {Rocco}}, \ and\ \bibinfo {author}
  {\bibfnamefont {R.~B.}\ \bibnamefont {Wiringa}},\ }\href {\doibase
  10.1103/PhysRevC.105.014002} {\bibfield  {journal} {\bibinfo  {journal}
  {Phys. Rev. C}\ }\textbf {\bibinfo {volume} {105}},\ \bibinfo {pages}
  {014002} (\bibinfo {year} {2022})},\ \Eprint
  {http://arxiv.org/abs/2108.10824} {arXiv:2108.10824 [nucl-th]} \BibitemShut
  {NoStop}%
\bibitem [{\citenamefont {Sobczyk}\ \emph {et~al.}(2020)\citenamefont
  {Sobczyk}, \citenamefont {Acharya}, \citenamefont {Bacca},\ and\
  \citenamefont {Hagen}}]{Sobczyk:2020qtw}%
  \BibitemOpen
  \bibfield  {author} {\bibinfo {author} {\bibfnamefont {J.~E.}\ \bibnamefont
  {Sobczyk}}, \bibinfo {author} {\bibfnamefont {B.}~\bibnamefont {Acharya}},
  \bibinfo {author} {\bibfnamefont {S.}~\bibnamefont {Bacca}}, \ and\ \bibinfo
  {author} {\bibfnamefont {G.}~\bibnamefont {Hagen}},\ }\href {\doibase
  10.1103/PhysRevC.102.064312} {\bibfield  {journal} {\bibinfo  {journal}
  {Phys. Rev. C}\ }\textbf {\bibinfo {volume} {102}},\ \bibinfo {pages}
  {064312} (\bibinfo {year} {2020})},\ \Eprint
  {http://arxiv.org/abs/2009.01761} {arXiv:2009.01761 [nucl-th]} \BibitemShut
  {NoStop}%
\bibitem [{\citenamefont {Sobczyk}\ \emph {et~al.}(2023)\citenamefont
  {Sobczyk}, \citenamefont {Acharya}, \citenamefont {Bacca},\ and\
  \citenamefont {Hagen}}]{Sobczyk:2023sxh}%
  \BibitemOpen
  \bibfield  {author} {\bibinfo {author} {\bibfnamefont {J.}~\bibnamefont
  {Sobczyk}}, \bibinfo {author} {\bibfnamefont {B.}~\bibnamefont {Acharya}},
  \bibinfo {author} {\bibfnamefont {S.}~\bibnamefont {Bacca}}, \ and\ \bibinfo
  {author} {\bibfnamefont {G.}~\bibnamefont {Hagen}},\ }\href@noop {} {\
  (\bibinfo {year} {2023})},\ \Eprint {http://arxiv.org/abs/2310.03109}
  {arXiv:2310.03109 [nucl-th]} \BibitemShut {NoStop}%
\bibitem [{\citenamefont {Bacca}\ \emph {et~al.}(2014)\citenamefont {Bacca},
  \citenamefont {Barnea}, \citenamefont {Hagen}, \citenamefont {Miorelli},
  \citenamefont {Orlandini},\ and\ \citenamefont {Papenbrock}}]{Bacca:2014rta}%
  \BibitemOpen
  \bibfield  {author} {\bibinfo {author} {\bibfnamefont {S.}~\bibnamefont
  {Bacca}}, \bibinfo {author} {\bibfnamefont {N.}~\bibnamefont {Barnea}},
  \bibinfo {author} {\bibfnamefont {G.}~\bibnamefont {Hagen}}, \bibinfo
  {author} {\bibfnamefont {M.}~\bibnamefont {Miorelli}}, \bibinfo {author}
  {\bibfnamefont {G.}~\bibnamefont {Orlandini}}, \ and\ \bibinfo {author}
  {\bibfnamefont {T.}~\bibnamefont {Papenbrock}},\ }\href {\doibase
  10.1103/PhysRevC.90.064619} {\bibfield  {journal} {\bibinfo  {journal} {Phys.
  Rev. C}\ }\textbf {\bibinfo {volume} {90}},\ \bibinfo {pages} {064619}
  (\bibinfo {year} {2014})},\ \Eprint {http://arxiv.org/abs/1410.2258}
  {arXiv:1410.2258 [nucl-th]} \BibitemShut {NoStop}%
\bibitem [{\citenamefont {Efros}\ \emph {et~al.}(1999)\citenamefont {Efros},
  \citenamefont {Leidemann},\ and\ \citenamefont {Orlandini}}]{Efros:1999ab}%
  \BibitemOpen
  \bibfield  {author} {\bibinfo {author} {\bibfnamefont {V.~D.}\ \bibnamefont
  {Efros}}, \bibinfo {author} {\bibfnamefont {W.}~\bibnamefont {Leidemann}}, \
  and\ \bibinfo {author} {\bibfnamefont {G.}~\bibnamefont {Orlandini}},\ }\href
  {\doibase 10.1007/s006010050118} {\bibfield  {journal} {\bibinfo  {journal}
  {Few Body Syst.}\ }\textbf {\bibinfo {volume} {26}},\ \bibinfo {pages} {251}
  (\bibinfo {year} {1999})}\BibitemShut {NoStop}%
\end{thebibliography}%

\end{document}